\documentclass[twocolumn]{aastex631}
\usepackage{physics}
\usepackage{placeins}
\usepackage{multirow}
\usepackage{subfigure}
\usepackage{afterpage}
\usepackage{stackengine}
\newcommand{\LAS}[1]{#1}

\begin{document}

\title{Time-Dependent Turbulent Electron Acceleration and Transport in Solar Flares}

\correspondingauthor{Luiz Schiavo}
\email{luiz.schiavo@northumbria.ac.uk}

\author[0000-0002-5082-1398]{Luiz A. C. A. Schiavo}
\affiliation{Northumbria University, Newcastle upon Tyne, NE1 8ST, UK}

\author[0000-0002-5082-1398]{Natasha L. S. Jeffrey}
\affiliation{Northumbria University, Newcastle upon Tyne, NE1 8ST, UK}

\author[0000-0002-5915-697X]{Gert J. J. Botha}
\affiliation{Northumbria University, Newcastle upon Tyne, NE1 8ST, UK}

\author[0000-0002-7863-624X]{James A. McLaughlin}
\affiliation{Northumbria University, Newcastle upon Tyne, NE1 8ST, UK}



 \begin{abstract}
Solar flares are explosive releases of magnetic energy stored in the solar corona, driven by magnetic reconnection. These events accelerate electrons, generating hard X-ray emissions and often display Quasi Periodic Pulsations (QPPs) across the energy spectra. However, the energy transfer process remains poorly constrained, with competing theories proposing different acceleration mechanisms. We investigate electron acceleration and transport in a flaring coronal loop by solving a time-dependent Fokker-Planck equation. Our model incorporates transient turbulent acceleration, simulating the effects of impulsive energy input to emulate the dynamics of time-dependent reconnection processes. We compute the density-weighted electron flux, a diagnostic directly comparable to observed X-ray emissions, across the energy and spatial domains from the corona to the chromosphere. 
We investigate different time-dependent functional forms of the turbulent acceleration, finding that the functional form of the acceleration source maintains its signature across energy bands (1 to 100 keV) with a response time that is energy dependent (with higher energy bands displaying a longer response time). In addition, we find that (a) for a square pulse the switch on and off response time is different; (b) for a sinusoidal input the periodicity is preserved; and (c) for a damped sinusoidal the decay rate increases with density and higher energy bands lose energy faster.
This work presents a novel methodology for analyzing electron acceleration and transport in flares driven by time-dependent sources.
 \end{abstract}



\section{Introduction} \label{sec:intro}
\typeout{Textwidth: \the\textwidth}
\typeout{Columnwidth: \the\columnwidth}
Solar flares represent one of the largest energy release processes in the solar system, converting stored magnetic energy into particle acceleration, plasma heating, and radiation through magnetic reconnection (\cite{parker_sweets_1957}).
These highly energetic events emit radiation spanning from radio waves to $\gamma$-rays and are closely linked to particle acceleration \citep{fletcher_observational_2011}.
These events generate nonthermal electron populations that produce observable hard X-ray (HXR) emissions, mainly electron-ion bremsstrahlung \citep{benz_flare_2008}, providing critical diagnostics of energy release and particle transport. Since the pioneering HXR observations of \cite{peterson_gamma-ray_1959} and \cite{frost_rapid_1969}, flare studies have revealed a complex interplay between electron acceleration and transport, with key insights coming from spatially resolved spectroscopy (e.g., Reuven Ramaty High Energy Solar Spectroscopic Imager, RHESSI: \cite{lin_reuven_2002}; Solar Orbiter/Spectrometer Telescope for Imaging X-rays, STIX: \cite{krucker_spectrometertelescope_2020}).

Observations show that flare-accelerated electrons produce both coronal loop top and chromospheric foot point HXR sources \citep{hoyng_origin_1981,masuda_loop-top_1994}, with spectral differences between these sources suggesting partial electron trapping in the corona \citep{petrosian_stochastic_2012,battaglia_relations_2006,simoes_implications_2013}. The standard thick-target model \citep{brown_deduction_1971} and its warm-target extension \citep{kontar_collisional_2015} provide a framework for electron transport but rely on steady-state assumptions, neglecting key transient aspects of flare energy release, such as quasi-periodic pulsations (QPPs). 

QPPs are oscillations that appear as transient modulations in flare emission, are frequently observed across multiple wavelengths, including microwave emissions \citep{nakariakov_quasi-periodic_2018}, extreme ultraviolet  \citep[e.g.][]{dominique_detection_2018,Li2025}, soft and hard X-rays \citep[e.g.][]{simoes_soft_2015,dennis_detection_2017,shi_multiwavelength_2024} and gamma-rays \citep{nakariakov_quasi-periodic_2010}. QPPs are a common observational feature in solar and stellar flare emissions, though their underlying mechanisms remain unclear \citep{McLaughlin2018, Kupriyanova2020-lp, Zimovets2021}. They are present during the impulsive and decaying phase of solar flares \citep[e.g.][]{hayes_quasi-periodic_2016,Collier2024}, with a typical period ranging from seconds to minutes. Recently observational studies point to evidence that they may originate around coronal loop tops and possibly by a time-dependent magnetic reconnection mechanism \citep{yuan_compact_2019,purkhart_spatiotemporal_2025}, but in contrast \cite{song_unveiling_2025} found a QPP where the oscillations were primarily originating from the flare ribbon. 
{Additionally, MHD simulations with test particles showed that reconnection can lead to pulsations in the microwave emission similar to observed flare quasi-periodic pulsations without an oscillatory external driver \citep{BROWNING2024100049}. }
A better understanding of the origins and nature of QPPs could shed light on the nature of energy release in flares, and allow the pulsations to be used as a seismological diagnostic of the physical conditions in the flaring plasma. 
%
%


\LAS{
There is growing evidence to suggest that turbulence plays a fundamental role in particle acceleration during flares  \citep{larosa_mechanism_1993,petrosian_stochastic_2012,klein_acceleration_2017}. Magnetic reconnection can also create turbulence: reconnection self-consistently generates turbulence at the sub-ion scales independently from the existence of a fully developed spectrum at MHD scales \citep{franci_2017}, as well as accelerate super-thermal electrons up to relativistic energies \citep{2020ApJ...894..136T}. Reconnection outflow jets and associated instabilities (e.g., the Kelvin-Helmholtz instability) also drive a cascade of turbulent structures down to electron scales \citep{che_2020, crawford_2024,crawford_2025} and within these structures, such as expanding magnetic vortices, particles can be stochastically accelerated by induced electric fields via a second-order Fermi process.}


Turbulence as well as plasma waves are identified as possible candidates for causing the broadening of optically thin lines in soft X-rays, EUV and UV \citep{milligan_extreme_2015,fletcher_solar_2024}.
Observations reveal turbulent motions throughout coronal loops \citep{stores_spatial_2021}, with particularly strong signatures near magnetic reconnection sites \citep{french_dynamics_2020}. These findings support models where turbulent fluctuations \LAS{at kinetic scales} both accelerate electrons through second-order Fermi processes and confine them through pitch-angle scattering \citep{musset_diffusive_2018}. 
Previous studies \citep{chen_determination_2013,stackhouse_spatially_2018,stores_spectral_2023} only consider a time-independent diffusion coefficient to generate turbulent acceleration, neglecting the impulsive and transient nature of energy release in flares.

In this study, we analyze the electron acceleration and transport in a flaring coronal loop under a transient driver placed at the loop top. 
We solve a transient Fokker-Planck equation which incorporates collisional energy losses, collisional pitch-angle scattering and turbulent acceleration.
Here we investigate pulses and/or wave trains, simulating the transient energy injection around the loop top that emulates a possible QPP driver (e.g., transient reconnection) as discussed by \cite{yuan_compact_2019,purkhart_spatiotemporal_2025}. The main objective is to understand how the transient driver signature manifests across the different energy bands.

We also added an electron replenishment mechanism that maintains the accelerated population over extended durations, overcoming a limitation of previous models, allowing the simulation to run for longer simulation periods. Thus, this work represents the time-dependent model of turbulent electron acceleration incorporating both impulsive energy injection and sustained particle replenishment.

The paper is organized as follows: \S\ref{sec:numerical-model} details our numerical approach; \S\ref{sec:results} presents the results, including the replenishing procedure (\S\ref{sec:repleninshing}), and the analysis of the system when accelerated by a square pulse (\S\ref{sec:square-pulse}), a sinusoidal wave (\S\ref{sec:sine-wave}), and a damped sinusoidal wave (\S\ref{sec:damped-wave}); with conclusions in \S\ref{sec:conclusions}. Appendix \ref{appendix-eq} presents a derivation of the Fokker-Planck equation coefficients and Appendix \ref{appendix-fitting} details the sigmoid fitting from Section \ref{section-3.2.2}.

\section{Numerical model}\label{sec:numerical-model}

\subsection{Governing Fokker-Planck equation}
To model the evolution of electron acceleration and transport in a flaring coronal loop, a time-dependent Fokker-Planck equation is used. It utilizes a simplification of a three-dimensional Fokker-Planck equation from \cite{lifshitz1995physical}, which is based on spherical coordinates and assumes azimuthal symmetry \citep{KARNEY1986}. This equation is given by
\begin{eqnarray}
\frac{\partial f}{\partial t} &+& \mu v \frac{\partial f}{\partial z}  = 
\underbrace{\frac{1}{v^2}\pdv{}{v}\left[v^2D_{\rm{turb}}\pdv{f}{v}\right]}_{\mbox{turbulent acceleration}} \nonumber\\
&+& \underbrace{\frac{\Gamma}{2v^2} \left[ \frac{\partial}{\partial v} \left( 2v G(u) \frac{\partial f}{\partial v} + 4u^2 G(u) f \right) \right]}_{\mbox{collisional energy losses}} \nonumber \\ 
&+& \underbrace{\frac{\Gamma}{2v^2} \left[ \frac{1}{v} \frac{\partial}{\partial \mu} \left( (1 - \mu^2) \left( \text{erf}(u) - G(u) \right) \frac{\partial f}{\partial \mu} \right) \right]}_{\mbox{collisional pitch-angle scattering}},
\label{eq:fp-original}    
\end{eqnarray}
\noindent where $f(z,v,\mu,t)$ is the phase-space distribution function (electrons cm$^{-3}$ (cm s$^{-1}$)$^{-3}$), $v$ is the particle velocity (cm s$^{-1}$), $\mu$ the cosine of the pitch angle to the guiding magnetic field, $t$ is time (s) and $z$ the spatial coordinate (cm). In Eq.\ (\ref{eq:fp-original}) $m_e$ is the electron mass ($g$), $u$ is a dimensionless velocity $u=v/(\sqrt{2}v_{th})$, where the thermal velocity (cm s$^{-1}$) is given by $v_{th}=\sqrt{k_bT/m_e}$, where $k_b$ is the Boltzmann constant (erg K$^{-1}$), $T$ is the temperature (K), and $e$ is the electron charge (esu). The function $G(u)$ is defined in terms of the error function erf and the error function derivative $\mbox{erf}'$ as
\begin{equation}
G(u) = \frac{\mbox{erf}(u)-u\:\mbox{erf}'(u)}{2u^2} 
\label{eq:G}
\end{equation}
\noindent and
\begin{equation}
\mbox{erf}(u) =\frac{2}{\sqrt{\pi}}\int_{0}^{u}\exp(-\zeta^2)d\zeta .
\end{equation}
These functions regulate the low-energy electron interactions, where $E \approx k_b T$, ensuring they become indistinguishable from the background thermal plasma \citep{jeffrey_variation_2014,stores_spectral_2023}. Finally, in Eq.\ (\ref{eq:fp-original}) $\Gamma$ is defined as $\Gamma=4\pi e^4 \ln{(\Lambda)} n/m_e^2$ (cm$^3$s$^{-4}$), where $\ln{(\Lambda)}$ represents the Coulomb logarithm, and $n$ the number density (cm$^{-3}$).

In our model, electrons can be accelerated over an extended region by a turbulent diffusion coefficient, as previously studied in \cite{stackhouse_spatially_2018} and \cite{stores_spectral_2023}. The turbulent acceleration diffusion coefficient, denoted as $D_{\text{turb}}$, is modeled as follows:
\begin{equation}
D_{\rm{turb}} = \frac{v_{th}^2}{\tau_{\rm{acc}}}\left(\frac{v}{v_{th}}\right)^\chi h(z) g(t) .
\label{eq:Dturb}
\end{equation}
In this model, we introduce a time-dependent amplitude, denoted as $g(t)$, which can be used for modeling a transient source. This differs from the approaches taken by \cite{stackhouse_spatially_2018} and \cite{stores_spectral_2023} where the diffusion coefficient is constant in time. Here, $\tau_{\rm{acc}}$ represents the selected acceleration timescales, $\chi$ is a constant that controls the velocity dependence, here chosen as $\chi=3$, and $h(z)$ describes a specific spatial distribution of turbulence in the coronal loop. Following \citet{stackhouse_spatially_2018,stores_spectral_2023}, the spatial distribution is modeled as a Gaussian function
\begin{equation}
h(z) = e^{-z^2/(2\sigma^2)} ,
\end{equation}
where $\sigma=3^{\prime\prime}$ is the size of the acceleration region. 

\begin{figure}[htb]
	\centering
	\includegraphics[width=0.99\columnwidth]{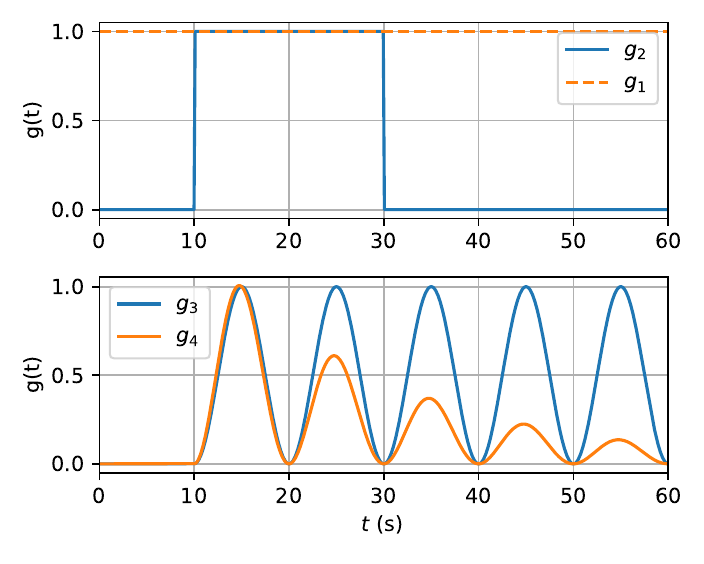}
	\caption{The evolution of the turbulent acceleration time amplitude, $g(t)$, described by Eqs.\ (\ref{eq:g2})-(\ref{eq:g4}). }
	\label{fig:g}
\end{figure}

In our study, four different time profiles are studied to model $g(t)$. The first case, $g(t) = g_1 = 1$, recovers the steady, but spatially dependent $D_{\rm{turb}}$ applied in \cite{stores_spectral_2023}. The second case, $g(t) = g_2$, represents a square pulse; $ g(t) = g_3 $ describes a sinusoidal wave; and $g(t) = g_4$ is a damped sinusoidal wave. The definition of $ g$ is provided in equations (\ref{eq:g2}) to (\ref{eq:g4}), and the profile of $g(t)$ is illustrated in Figure \ref{fig:g}. The transient pulse and wave trains are defined as
\begin{eqnarray}
g_2(t) &=& 
\begin{cases}
1, & \text{for } 10\  \text{s} < t <30 \ \text{s} ,\\
0, & \mbox{otherwise} ,
\end{cases} \label{eq:g2} \\
g_3(t) &=& H(t-t_{\rm{start}}) \sin^2\left[ \frac{2\pi}{\lambda}(t-t_{\rm{start}}) \right] , \label{eq:g3}\\
g_4(t) &=& g_3 e^{-(t-15)/20} ,  \label{eq:g4}
\end{eqnarray}
\noindent where $\lambda=20$ s, $t_{\rm{start}}=10$~s is when the acceleration starts, and $H$ is the Heaviside function. Note that the actual oscillation period for $g_3$ and $g_4$ is $\lambda/2$, since $2\sin^2x = 1-\cos2x$.
\LAS{Note that $g(t)$ is a phenomenological parameterization of the impulsive energy input into the turbulent cascade. The functional forms we explore (square, sinusoidal, damped sinusoidal) are not intended to perfectly model the kinetic linear growth and saturation of a specific instability but are chosen as simple, well-defined impulses to probe the system's response characteristics.}

\subsection{Stochastic Differential Equations}
The Fokker-Planck equation, Eq. (\ref{eq:fp-original}), is converted to an equivalent system of stochastic differential equations (SDE) in It$\hat{\rm{o}}$ form \citep{gardiner1985handbook}
\begin{eqnarray}
    v_{n+1} &=& v_{n} +  (A_v+A_{vs}) \Delta t + \sqrt{(B_{v}+B_{vs})\Delta t} \ W_v ,\hspace{5mm} \label{eq:v} \\
    \mu_{n+1} &=& \mu_{n} + A_\mu \Delta t + \sqrt{B_{\mu}\Delta t}\:W_\mu , \label{eq:mu} \\
    z_{n+1} &=& z_{n}  + A_z \Delta t \label{eq:z} ,
\end{eqnarray}
where $W_v$ and $W_\mu$ are the Wiener process, which follow a Gaussian distribution with zero mean and unitary variance, and $\Delta t$ is the simulation time step. The other coefficients are defined as
\begin{eqnarray}%
&A_{z}  &=  \mu v , \\
&A_{v} &=  -\frac{\Gamma}{v^2}\left( \mbox{erf}(u) - 2\mbox{erf}'(u) + G(u) \right), \\
&B_{v} &= \frac{2\Gamma  G(u)}{v} , \\
&A_{vs} &= \frac{2D_{\rm{turb}}}{v}+\pdv{D_{\rm{turb}}}{v}  ,\\
&B_{vs} &=  2D_{\rm{turb}} , \\
&A_{\mu} &= -\frac{ \Gamma \mu}{v^3} \left( \text{erf}(u) - G(u) \right) , \\
&B_{\mu} &= \frac{\Gamma}{v^3}(1 - \mu^2) \left( \text{erf}(u) - G(u) \right) .
\end{eqnarray}
A complete description of the derivation is available in Appendix \ref{appendix-eq}.

\subsection{Plasma and boundary conditions}
\subsubsection{Loop model}
\begin{figure}[htb]
	\centering
	\includegraphics[trim = 10 0 10 5, width=0.99\columnwidth]{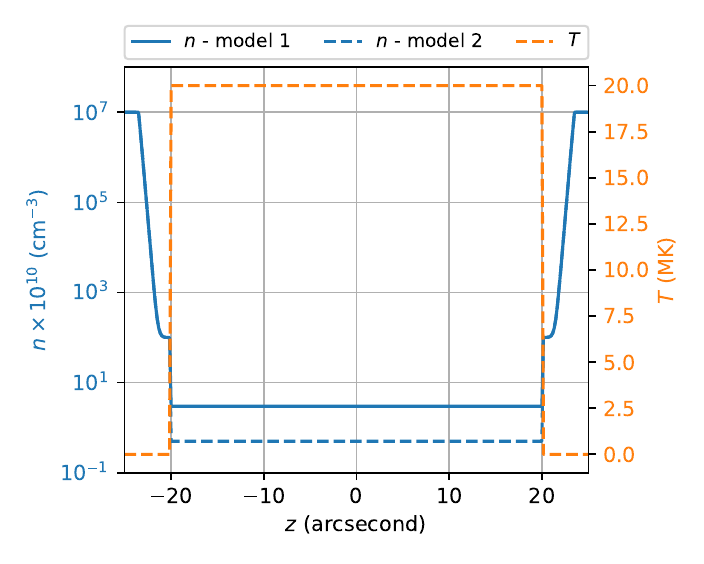}
	\caption{The two atmospheric models used in the simulations, showing the plasma number density (blue) and temperature distribution (orange). In the coronal region, the continuous blue line represents atmospheric model 1 and dashed blue atmospheric model 2. Outside the corona models 1 and 2 are identical.}
	\label{fig:atmosphere}
\end{figure}

Two loop models are examined (Figure \ref{fig:atmosphere}). 
Each loop is symmetric, with the loop apex placed at $z=0^{\prime\prime}$, the total length of the coronal region is $40^{\prime\prime}$ and the chromosphere starts at $|z| \pm 20^{\prime\prime}$. In the coronal region, the plasma temperature is 20 MK with $ \ln(\Lambda) = 20 $. Atmospheric model 1 uses a plasma number density of $ n_{\text{corona}} = 3 \times 10^{10} $ cm$^{-3}$ in corona. In contrast, atmospheric model 2 uses a lower plasma number density of $n_{\text{corona}} = 5 \times 10^{9} $ cm$^{-3}$ in the corona. 
Both atmospheric models feature exponential growth in the plasma number density inside the chromosphere, which is located at $20^{\prime\prime}<|z| < 23.5^{\prime\prime}$. The plasma number density increases to $10^{12}$ cm$^{-3}$ at the top of the chromosphere and then rises exponentially towards the photosphere (which is at $z = 23.5^{\prime\prime}$), at which point it attains the value $10^{17}$cm$^{-3}$. Thus, the plasma number density (as shown in Fig.\ \ref{fig:atmosphere}) is described by
\begin{eqnarray*}
n &=& n_{\text{corona}} \text{ for }  |z| < 20^{\prime\prime} , \\
n &=& n_{\text{chromo}} + n_{\text{photo}} e^{\left(\frac{|z| -23.5^{\prime\prime}}{h_0}\right)}  \text{ for }  20^{\prime\prime}\le |z| < 23.5^{\prime\prime}, 
\end{eqnarray*}
where $n_{\text{chromo}} = 10^{12}$cm$^{-3}$ and $n_{\text{photo}} = 1.16 \times 10^{17}$cm$^{-3}$. For $|z|\geq 23.5^{\prime\prime}$ the number density is kept constant. The scale height, $h_0$, is assumed as $h_0$=0.18$^{\prime\prime}$ following \citet{battaglia_numerical_2012,jeffrey_role_2019}.
For both atmospheric models, the temperature in the chromosphere is considered to be a cold target (achieved using $ T = 0.01 $ K) and $\ln(\Lambda) = 7 $. 
This temperature does not match the actual solar chromosphere temperature but guarantees that thermal motion is minimized and electron energy losses by advection dominate. Atmospheric model 1 is identical to \cite{jeffrey_role_2019, stores_spectral_2023} and atmospheric model 2 was chosen to investigate the effect of the plasma number density on electron transport and acceleration.

\subsubsection{Low velocity and energy limits}
For $v << v_{th}$, the SDEs diverge as $v \rightarrow 0$. In this limit, the functions of erf$(u)$ and its derivative and $G(u)$ can be approximated using their Maclaurin series expansion. The implementation follows \cite{pallister2023} where in the coronal region it is applied as a low velocity boundary condition: if
\begin{equation}
    v_n \leq \left( \Gamma \frac{8}{3} \sqrt{\frac{m_e}{2\pi k_B T} \Delta t} \right)^{1/2} = v_{\rm{limit}},
\end{equation}
\noindent the velocity at the next iteration is obtained by
\begin{equation}
v_{n+1} = \left( v_n^2 + v_{\rm{limit}}^2 \right)^{1/2} .
\end{equation}
In this implementation $\mu$ is randomly drawn from an isotropic distribution within the range $\mu = -1$ to $\mu = 1$. In the chromosphere, electrons that reach the region where $|z| \ge  20^{\prime\prime}$ and have the energy either below a certain threshold ($E < E_{min} = 0.1$ keV) or above another threshold ($E > E_{max} = 500$ keV) are removed from the simulation. This energy boundary condition is only applied when the electrons enter the chromosphere, and such electrons are considered lost to the sun. This approach allows the simulation to be completed in a reasonable time frame, as the collisional timescales for electrons in the chromosphere are orders of magnitude smaller than those in the corona.

\subsubsection{Electron replenishing condition}

The time that one injected thermal electron takes to leave the corona and deposit its energy in the chromosphere, is dominated by the diffusion time scale, $\tau_D$, and can be estimated analytically by
\begin{equation}
    \tau_D = \frac{\Gamma L^2}{12\sqrt{\pi}}\left(\frac{2m_e}{k_BT}\right)^{5/2} ,
\label{eq:taud}
\end{equation}
where $L$ is the half-loop length. The time scales for a thermal electron to deposit the energy at the foot of the loop for atmospheric models 1 and 2 are $\tau_D\approx17$ s and $\tau_D=3$ s, while higher energy electrons ($E>50$ keV) aligned with the magnetic field will leave the chromosphere on sub-second timescales. These fast deposit time scales create a limitation for running the simulation for a longer period than $\tau_{D}$.

To enable the simulations to run longer, so that time-dependent acceleration can be studied, an electron replenishing condition was added to every simulation. We define a region $|z|<z_{\text{repl}}$ where $z_{\text{repl}}=3^{\prime\prime}$, as the size of the region where electron population is kept constant.
During each simulation time step, we determine the number of electrons that left the replenishing region, $|z|<z_{\text{repl}}$, and inject the same number back into the region. 
The replenished electrons are drawn from a thermal ($T=20$~MK) distribution, with energies ranging between 1 keV and 20 keV. Additionally, the angular distribution of these thermal electrons is assumed to be isotropic, characterized by a uniform $\mu$ distribution. 

To investigate the influence of the chosen spatial distribution of replenished electrons, we considered four distinct cases for the $ z$ distribution of the introduced electrons. In the first three cases, electrons are randomly added within the loop top region, with the replenishing region of sizes $|z|<1^{\prime\prime}$, $2^{\prime\prime}$, and $3^{\prime\prime}$, respectively. In the fourth case, the electrons are distributed according to a Gaussian profile centered at $z = 0$ with a variance of $3''$. These variations allow us to explore the sensitivity of the simulation results to the replenishing mechanism's spatial characteristics. This is discussed in section \ref{sec:repleninshing}.

Implementing this replenishing boundary condition extends the simulation time scales significantly, enabling a more comprehensive analysis of electron dynamics and energy transport in solar loops. This approach provides a robust framework for studying long-term particle behavior and its implications for solar atmospheric heating and particle acceleration processes. While our replenishing condition does not represent any physical model,
electrons in flares may be physically replenished by e.g., plasma motion and/or return currents \citep{emslie_effect_1980,diakonov_thermal_1988,zharkova_effect_2006,alaoui_understanding_2017} bringing electrons back to the acceleration region.

\subsection{Simulation setup}
We conducted a parametric study to assess the influence of a time-dependent turbulent diffusion coefficient, Eq.\ (\ref{eq:Dturb}).
We investigated the influence of the turbulent acceleration time scale, $\tau_{\rm{acc}}$, and different time profiles, $g(t)$, that are displayed in Fig.\ \ref{fig:g}. Table~\ref{tab:setup} summarizes the setup for each simulation run.
\begin{table*}[]
\centering
\begin{tabular}{lcccccc}\hline \hline
Case & $n$ (cm$^{-3}$) & $\tau_{\rm{acc}}$ (s) & $\tau_c$ (s) & $\tau_{\rm{acc}}/\tau_{c}$  & $g(t)$ & replenishing z\\ \hline
1  & $3\times 10^{10}$ & 10 & 0.011 & 915  & $g_1$ & Random at $|z|<1^{\prime\prime}$ \\
2  & $3\times 10^{10}$ & 10 & 0.011 & 915  & $g_1$ & Random at $|z|<2^{\prime\prime}$ \\
3  & $3\times 10^{10}$ & 10 & 0.011 & 915  & $g_1$ & Random at $|z|<3^{\prime\prime}$ \\
4  & $3\times 10^{10}$ & 10 & 0.011 & 915  & $g_1$ &  Gaussian with $\sigma=3^{\prime\prime}$  \\ \hline
5  & $3\times 10^{10}$ & 5  & 0.011 & 457  & $g_2$ & Random at $|z|<3^{\prime\prime}$ \\
6  & $3\times 10^{10}$ & 10 & 0.011 & 915  & $g_2$ & Random at $|z|<3^{\prime\prime}$ \\
7  & $3\times 10^{10}$ & 15 & 0.011 & 1372 & $g_2$ & Random at $|z|<3^{\prime\prime}$ \\
8  & $3\times 10^{10}$ & 20 & 0.011 & 1829 & $g_2$ & Random at $|z|<3^{\prime\prime}$ \\ 
9  & $3\times 10^{10}$ & 25 & 0.011 & 2286 & $g_2$ & Random at $|z|<3^{\prime\prime}$ \\ \hline
10 & $5\times 10^9$    & 5  & 0.066 & 76   & $g_2$ & Random at $|z|<3^{\prime\prime}$ \\
11 & $5\times 10^9$    & 10 & 0.066 & 152  & $g_2$ & Random at $|z|<3^{\prime\prime}$ \\
12 & $5\times 10^9$    & 15 & 0.066 & 229  & $g_2$ & Random at $|z|<3^{\prime\prime}$ \\
13 & $5\times 10^9$    & 20 & 0.066 & 305  & $g_2$ & Random at $|z|<3^{\prime\prime}$ \\
14 & $5\times 10^9$    & 25 & 0.066 & 381  & $g_2$ & Random at $|z|<3^{\prime\prime}$ \\ \hline
15 & $3\times 10^{10}$ & 5  & 0.011 & 457  & $g_3$ & Random at $|z|<3^{\prime\prime}$ \\
16 & $3\times 10^{10}$ & 25 & 0.011 & 2286 & $g_3$ & Random at $|z|<3^{\prime\prime}$ \\ \hline
17 & $5\times 10^9$    & 5  & 0.066 & 76   & $g_3$ & Random at $|z|<3^{\prime\prime}$ \\
18 & $5\times 10^9$    & 25 & 0.066 & 381  & $g_3$ & Random at $|z|<3^{\prime\prime}$ \\ \hline
19 & $3\times 10^{10}$    & 5  & 0.011 & 83   & $g_4$ & Random at $|z|<3^{\prime\prime}$ \\
20 & $3\times 10^{10}$    & 25 & 0.011 & 381  & $g_4$ & Random at $|z|<3^{\prime\prime}$ \\ \hline
21 & $5\times 10^9$    & 5  & 0.066 & 83   & $g_4$ & Random at $|z|<3^{\prime\prime}$ \\
22 & $5\times 10^9$    & 25 & 0.066 & 381  & $g_4$ & Random at $|z|<3^{\prime\prime}$ \\ \hline
\end{tabular}
\centering
	\caption{The simulation setup for each run, containing plasma number density in the corona region, $n$, turbulent acceleration time scale, $\tau_{\rm{acc}}$, collisional time scale, $\tau_c$, the acceleration term time profile, $g(t)$, and the replenishing profile along the loop. }
    \label{tab:setup}
\end{table*}
Simulation sets 1 to 4 utilize a steady source term ($g_1$) to analyze the effect of the electron replenishing on the observed electron energy spectra. These simulations provide insights into how the replenishing mechanism affects the temporal evolution of electron populations and their corresponding radiative signatures.

Simulation sets 5 to 9 examine the impact of the turbulent acceleration time scale with a square pulse, $g_2$. We investigated how different acceleration regimes influence the electron energy distribution by varying the turbulent acceleration parameters. The simulation sets 10 to 14 investigate the response of the same square pulse under a different atmospheric condition. Specifically, these simulations use atmospheric model 2, where the collisional time scale, $ \tau_{c} \approx v_{th}^3 / \Gamma $, is six times larger than in atmospheric model 1 due to the density decrease. This allows us to study how changes in collisionality affect electron acceleration and transport properties.

Simulation sets 15 and 18 test the electron response to a wave source. These simulations explore the effects of periodic perturbations on electron transport and the resultant electron distribution. Finally, 19 to 22 analyze the influence of a damped wave for the electron acceleration and transport.

A time step of $\Delta t = 10^{-3} $ s is used for every simulation, ensuring that it is smaller than the collisional and turbulent acceleration times. This choice of time step guarantees numerical stability and accurate resolution of the physical processes under investigation. We also tested case 3 with a smaller time step, $\Delta t = 10^{-4} $ s, which lead to simulation results similar to $\Delta t = 10^{-3} $ s.

All simulations have an initial population of 100,000 electrons. As electrons leave the replenishing region, additional electrons are introduced to maintain a constant population within this region defined by $|z| < z_{\text{repl}}$. After approximately 3 s, the simulation reaches an equilibrium condition, stabilizing at around 300,000 electrons. This equilibrium reflects a balance between electron losses and the replenishing mechanism, enabling long-term electron transport and acceleration simulations. Note that the initial number of electrons in the system is not as important as the number of electrons after the initial transient, when the number of electrons is stabilized in the system.

\subsection{Density-weighted electron energy spectrum, $nVF$}
The simulation outputs are plotted as the density-weighted electron energy distribution, $nVF$.
$nVF$ gives a measure of how turbulent acceleration affects observable electron properties, following the methodology established in previous studies \citep{brown_determination_2003}. This quantity provides a direct connection to X-ray observations while remaining independent of specific assumptions about coronal transport processes. For the angle-integrated case, the X-ray photon spectrum $ I(\epsilon) $ relates to $ nVF $ through the integral expression
\begin{equation}
I(\epsilon) = \frac{1}{4\pi R^2} \int_{\epsilon}^{\infty} nVF(E) \, Q(E, \epsilon) \, dE ,
\end{equation}
where $\epsilon$ represents the photon energy, $ R $ denotes the sun-observer distance, and $ Q(E, \epsilon) $ is the angle-integrated bremsstrahlung cross section \citep{koch_bremsstrahlung_1959}.

Within the framework of a basic thick-target model, the spectral index $\delta$ of the injected electron distribution typically relates to the observed $nVF$ spectral index $\delta_{nVF}$ through the approximate relation $\delta \approx \delta_{nVF} + 2$ as discussed in \cite{stores_spectral_2023}. Similarly, the resulting X-ray photon index $\gamma$ follows $\gamma \approx \delta_{nVF} + 1$ (e.g. \citet{battaglia_relations_2006,chen_impulsive_2012}). However, these relationships represent idealized cases, as numerous studies have demonstrated that various transport effects can significantly modify these index differences, either increasing or decreasing them depending on the specific physical conditions.
Since $nVF$ is a proxy of X-ray light curves, it allows us to study the time evolution of the system matching flare time scales. 
In solar flares, the X-ray spectra typically exhibit steep power-law distributions, with the spectral index serving as a key diagnostic parameter for characterizing the accelerated electron population. Unlike conventional thick-target models that assume pre-accelerated electron injection, our approach models the direct acceleration of electrons from a thermal distribution, providing a more self-consistent treatment of the particle acceleration process.

\section{Results}  \label{sec:results}

\subsection{Influence of the electron replenishing}  \label{sec:repleninshing}

Before analyzing the cases where the turbulent acceleration is time dependent, the effect of the replenishing is investigated on the electron spectra in cases with a time independent turbulent acceleration. Here, simulation cases 1 to 4 from Table \ref{tab:setup} are compared.
In cases 1 to 3, electrons are added randomly inside the replenishing zone, while in case 4 they are injected using a Gaussian distribution. The electrons follow a thermal energy distribution and have isotropic $\mu$ in all cases, using a coronal flaring temperature of 20 MK.
\begin{figure}[htb]
	\centering
	\includegraphics[width=0.94\columnwidth]{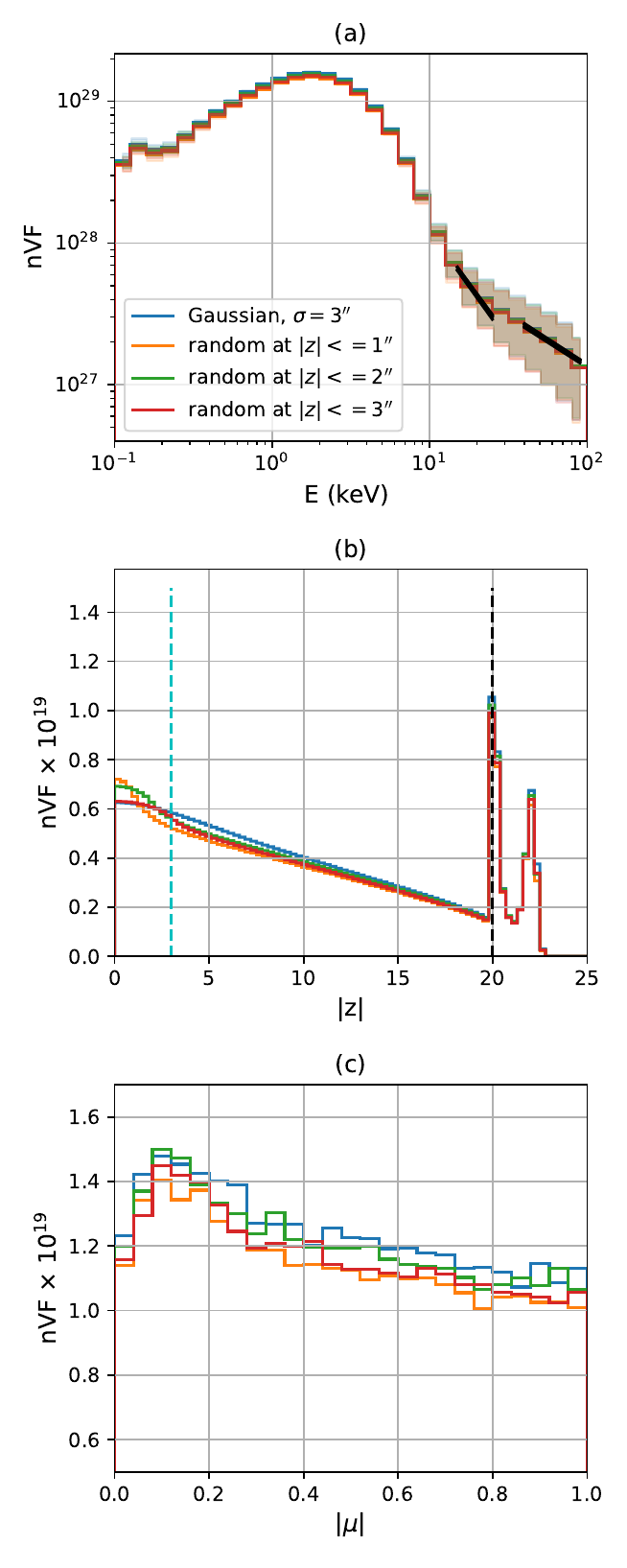}
	\caption{$nVF$ spectra as function of (a) the electron energy, the shaded areas represent the estimated uncertainty multiplied by 100 for better visualization, (b) loop position, and (c) pitch angle cosine. In panel (b) the black-dashed line represents the chromosphere boundary and the cyan-dashed line is the acceleration region, $\sigma=3^{\prime\prime}$.}
	\label{fig:replenish}
\end{figure}
Figure \ref{fig:replenish} presents the density-weighted spectra, $nVF$, for simulations cases 1-4 as function of (\ref{fig:replenish}a) energy, $E$, (\ref{fig:replenish}b) spatial distribution, $|z|$, and (\ref{fig:replenish}c) pitch angle cosine, $\mu$. Figure \ref{fig:replenish}a shows the formation of the non-thermal tail in the energy spectra, which is caused by turbulent acceleration consistent with \citet{stores_spectral_2023}. There is no significant difference in the energy distribution spectra based on the spatial distribution of the replenished electrons. 
The shaded regions indicate the estimated uncertainty for each histogram bin, computed as $\sqrt{\Sigma_i w_i^2}$, where $w_i$ denotes the weight of the $i$-th entry in the histogram bin, in this case $w_i$ is the number density. 
Figure \ref{fig:replenish}b shows the electron spatial distribution along the loop. In every case there is a deposition of electrons at the chromospheric boundary, $|z|=20^{\prime\prime}$, and a secondary peak around $|z|\approx22^{\prime\prime}$. In the acceleration region, $|z|<3^{\prime\prime}$, there is a larger difference between the simulations with the Gaussian and with random replenishing, which generates a slightly different slope in the $|z|$ distribution. The $nVF$ distributions in Figure \ref{fig:replenish}b reveal only minor differences at the loop apex while the chromosphere remains unchanged.

Figure \ref{fig:replenish}c displays a comparison of $nVF$ distributions as function of $|\mu|$ for cases 1-4, and once again, there is no significant difference. This finding provides us with confidence to analyze long simulation periods, as the electron replenishing process has a negligible effect on the results on the integrated $nVF$. Although the electrons were replenished using an isotropic distribution, the integrated $nVF$ does not remain isotropic. This indicates that some low pitch-angle electrons are being trapped creating a peak near $|\mu|=$0.1\footnote{This distribution is consistent with the results of \citet{stores_spectral_2023} where inefficient scattering can lead to trapping.}.

\subsection{Response to a square pulse}  \label{sec:square-pulse}
\subsubsection{Energy spectra}

\begin{figure*}[p]
	\centering
    	\includegraphics[width=0.99\textwidth]{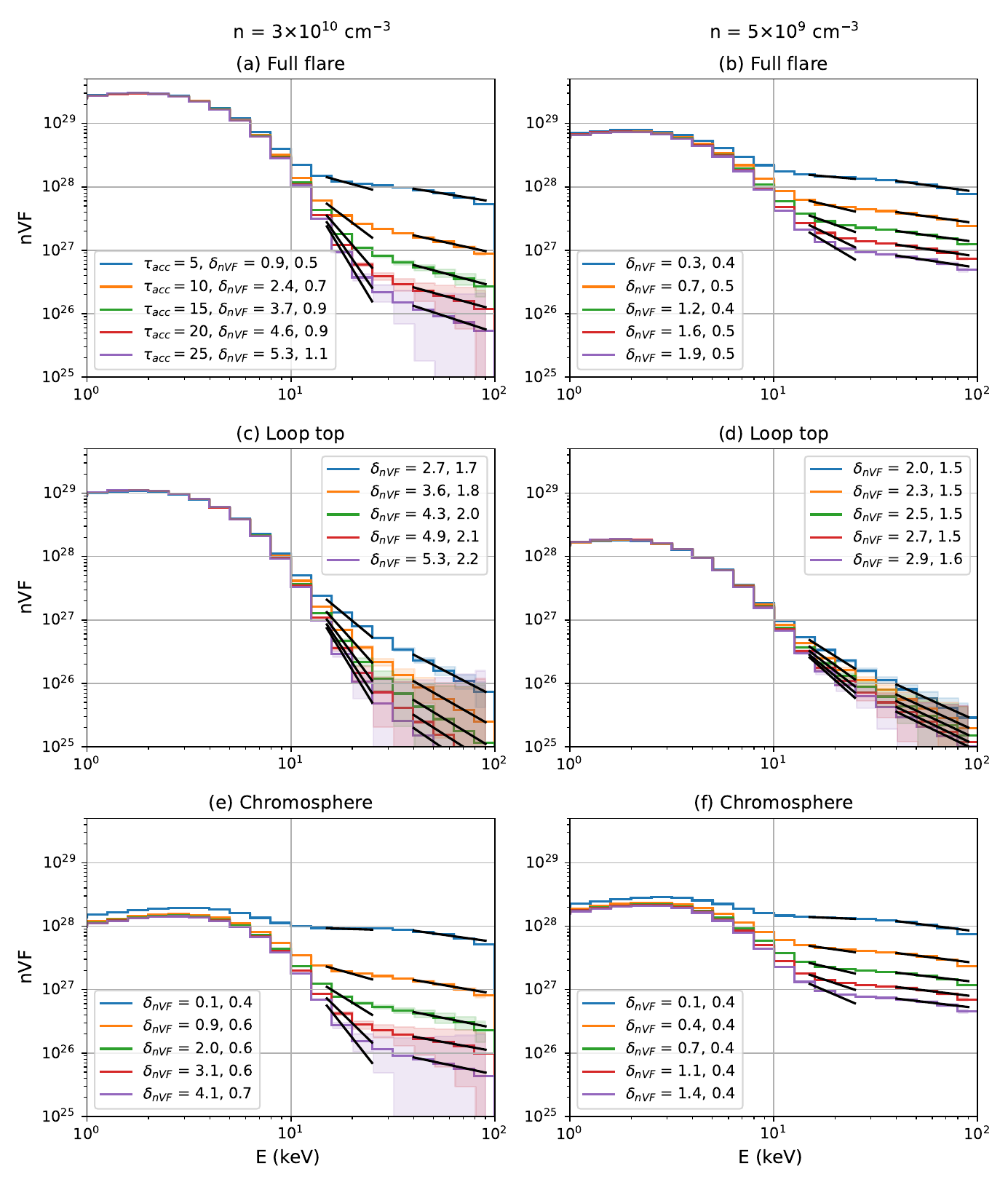}
	\caption{$nVF$ energy spectra for different regions of the solar flare for source $g_2$ (described in Eq.\ (\ref{eq:g2})). Left column panels (a), (c) and (e), contain results for $n=3\times 10^{10}$ cm$^{-3}$, while right column panels (b), (d) and (f) for $n=5\times 10^{9}$ cm$^{-3}$. The shaded regions represent the estimated uncertainty of each bin, multiplied by 100 for better visualization. $\tau_{\rm{acc}}$ color code in panel (a) stays the same across every panel. $\delta_{nVF}$ represents the spectral indices obtained applying a fitting on $nVF$ between 15-25 keV and 40-90 keV.}
	\label{fig:square-pulse-energy}
\end{figure*}

The analysis now focuses on transient turbulent acceleration cases which involve a square wave pulse, model $g_2$ described in Eq.\ (\ref{eq:g2}). The turbulent acceleration starts at $t=$ 10 s and accelerates electrons for 20 s, finishing abruptly. Figure \ref{fig:square-pulse-energy} shows the $nVF$ plotted as a function of electron energy for cases 5 to 9 and 10 to 14 in Table \ref{tab:setup}. The energy spectra were integrated from $t = 5$ seconds to $t = 60$ s by summing the energy bins obtained at each simulation time step. For all panels in Figure \ref{fig:square-pulse-energy} the cases with lower $\tau_{\rm{acc}}$ deviate more from the background plasma's thermal distribution.

In the left column section of Figure \ref{fig:square-pulse-energy} (Figures \ref{fig:square-pulse-energy}a, \ref{fig:square-pulse-energy}c and \ref{fig:square-pulse-energy}e), the high-density cases are presented, cases 5 to 9 from Table \ref{tab:setup}. A power-law fit (black solid line) determines the spectral index, $\delta_{nVF}$, between 40 to 90 keV and 15 to 25 keV. In the full flare ($\forall z$) energy spectra (Figure \ref{fig:square-pulse-energy}a) we observe the formation of a non-thermal tail, between 40-90 keV, with a spectral index $\delta_{nVF}$ ranging from 0.5 for $\tau_{\rm{acc}} = 5$ s to 1.1 s for $\tau_{\rm{acc}} = 25$ s in the 40-90 keV range. The non-thermal tail gradually decreases as $\tau_{\rm{acc}}$ increases. In the loop-top analysis (Figure \ref{fig:square-pulse-energy}c) we note a small variation in the spectral index across the simulated cases, ranging from 1.7 to 2.2 at 40-90 keV, with the non-thermal tail starting around 10 keV.

In the chromosphere ($|z| > 20^{\prime\prime}$) shown in Figure \ref{fig:square-pulse-energy}e, the variation in the spectral index for the non-thermal population is also small, ranging from 0.4 to 0.7 in the energy range of 40-90 keV. 
Analyzing the full flare spectra, it is seen there are significant changes in the energy range 40-90 keV mainly arising from the differences in the chromospheric energy signatures.

The lower plasma number density cases are shown in the right column of Figure \ref{fig:square-pulse-energy} (Figures \ref{fig:square-pulse-energy}b,  \ref{fig:square-pulse-energy}d and  \ref{fig:square-pulse-energy}f). These cases have a collisional time scale, $\tau_{c}$, six times larger than cases where number density is higher. In these lower number-density cases, negligible variation in the full flare, loop top, and chromosphere spectral indices between 40-90 keV are observed.

The acceleration time scale, $\tau_{\rm{acc}}$, significantly influences the full flare energy spectra as reported in \cite{stores_spectral_2023}. The spectral index of the full flare rises as $\tau_{\rm{acc}}$ increases, reducing the spectra at high energies, 40-90 keV, as well at 15 to 25 keV.  A larger variation is observed in the spectral index for the higher-density cases, changing from 0.9 to 5.3 in the energy range of 15 to 30 keV. In contrast, the lower density cases vary from 0.3 to 1.9 within the same energy range. In the loop-top region, where $|z| < 5^{\prime\prime}$, there is 6\% variation in the spectral index variation for lower number-density cases, while it is moderate for higher number-density cases, 30\% between 40-90 keV.

Although $g_2$ is different from $g_1$ shown in Figure \ref{fig:g}, the results observed agree with trends seen by \citet{stores_spectral_2023}, where the increasing of $\tau_{\rm{acc}}$ decreases $\delta_{nVF}$.

\subsubsection{Transient evolution of $nVF$ energy spectra }\label{section-3.2.2}

\begin{figure*}[p]
	\centering
	\includegraphics[width=0.99\textwidth]{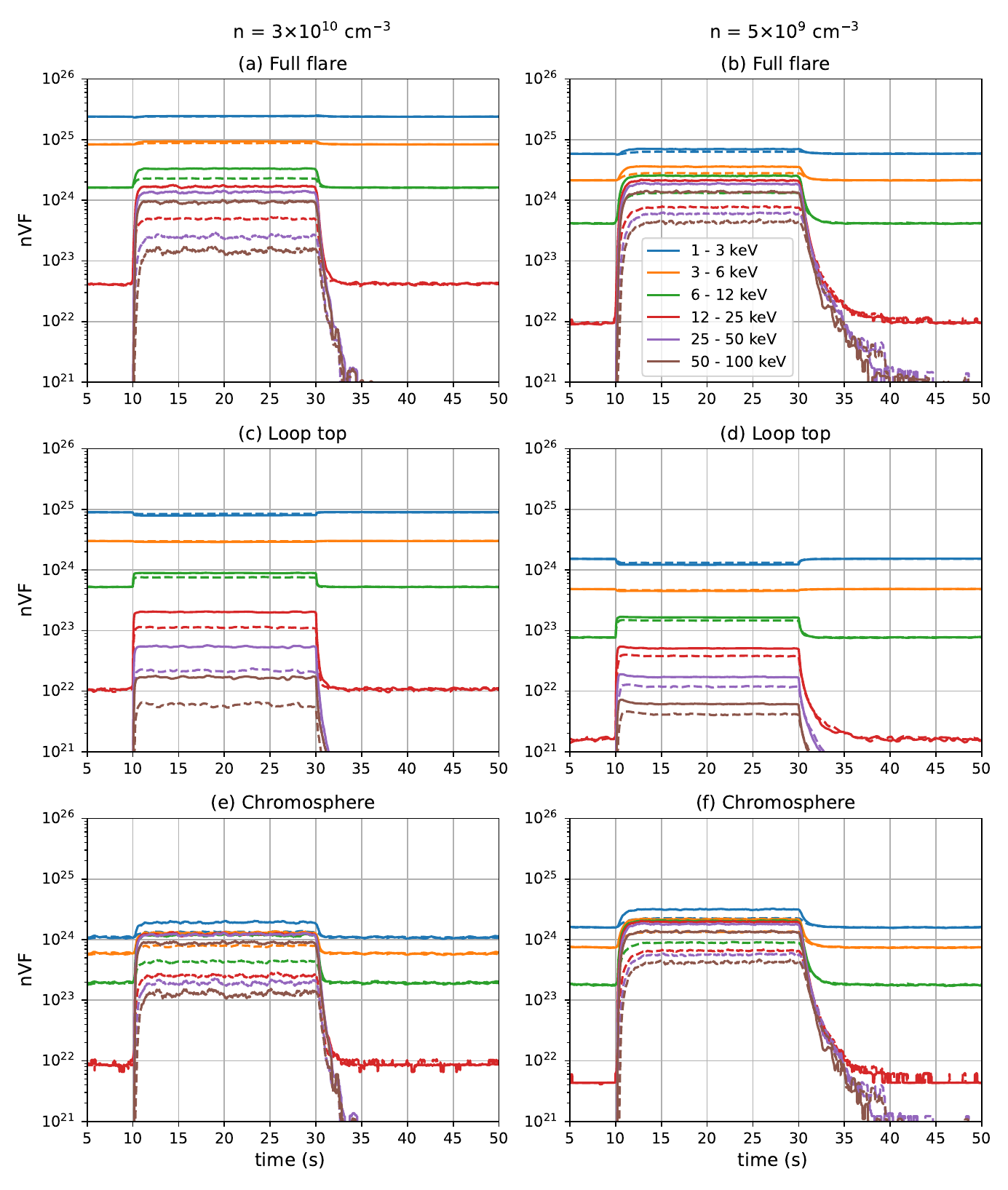}
	\caption{$nVF$ at different energy bands for a square pulse (source $g_2$, Eq.\ (\ref{eq:g2})), where turbulent acceleration turns on at $t=$ 10 s and turns off at $t=$ 30 s. Continuous lines show results for $\tau_{\rm{acc}}=$ 5 s, dashed lines show results for $\tau_{\rm{acc}}=$ 25 s. Left column, panels (a), (c) and (e), contain results for $n=3\times 10^{10}$ cm$^{-3}$, while right column, panels (b), (d) and (f) contain results for $n=5\times 10^{9}$ cm$^{-3}$.}
	\label{fig:energy-band-hat-tau5}
\end{figure*}

Figure \ref{fig:energy-band-hat-tau5} displays the temporal evolution of $nVF$ for cases 5, 9, 10, and 14 from Table \ref{tab:setup}, comparing acceleration timescales of $\tau_{\rm{acc}}=$ 5 using continuous lines and 25 s using dashed lines. The left column (Figures \ref{fig:energy-band-hat-tau5}a,  \ref{fig:energy-band-hat-tau5}c and  \ref{fig:energy-band-hat-tau5}e) contain results for high number-density conditions, while the right panels (Figures \ref{fig:energy-band-hat-tau5}b,  \ref{fig:energy-band-hat-tau5}d and  \ref{fig:energy-band-hat-tau5}f) show low number-density simulations. All spectra were integrated over 4 s period to provide enough statistics for each source implemented.

The high number-density cases (left panels) reveal distinct dynamical behavior across three regions: (\ref{fig:energy-band-hat-tau5}a) full flare, (\ref{fig:energy-band-hat-tau5}c) loop top, and (\ref{fig:energy-band-hat-tau5}e) chromosphere. A rapid $nVF$ increase occurs at $t=10$ s when the acceleration source starts, reaching a new steady state within 2 s. When acceleration terminates at $t=30$ s, $nVF$ returns to pre-acceleration levels within similar timescales. The loop-top region exhibits depletion in the 1-3 keV band, corresponding to electron acceleration that enhances $nVF$ at higher energy bands; this depletion coincides with increased chromospheric emission in the same energy range, demonstrating how the accelerated electrons gradually lose energy as they penetrate the denser chromospheric layers.

In contrast, the low number-density cases (Figures \ref{fig:energy-band-hat-tau5}b,  \ref{fig:energy-band-hat-tau5}d and  \ref{fig:energy-band-hat-tau5}f) show reduced amplitude variations between $\tau_{\rm{acc}}=5$ and 25 s compared to high number-density conditions. All cases exhibit characteristic relaxation timescales when acceleration initiates ($t=10$ s), with higher energy bands displaying longer relaxation time than energy bands. The $\tau_{\rm{acc}}=5$ s cases respond faster than $\tau_{\rm{acc}}=25$ s counterparts Figure \ref{fig:energy-band-hat-tau5}. Following acceleration cessation ($t=30$ s), the system requires extended recovery times to return to equilibrium, particularly in low number-density conditions where the increase of collisional time scale increases the electron transit times through the loop \citep{jeffrey_role_2019}.

To quantify the electron population's response to the square-wave acceleration pulse, the normalized $nVF$ signal (scaled to 0--1)  is fitted with a sigmoid function at the transitions at $t=10$ s (acceleration onset) and $t=30$ s (acceleration termination). The sigmoid function is defined as
\begin{equation}\label{eq:sigmoid}
S(t) = \frac{1}{1 + e^{-(t + t_s)\lambda_s}},
\end{equation}
where $t_s$ represents a time shift and $\lambda_s$ determines the growth rate. It characterizes the response timescale of the interval required for the electron population to transition from 5\% to 95\% of the full $nVF$ amplitude variation. This defines two distinct timescales: $\tau_{\rm{on}}$ (acceleration onset response) and $\tau_{\rm{off}}$ (acceleration termination response). 
\begin{figure}[htb]
	\centering
    \includegraphics[trim= 10 0 10  0 ,width=.99\columnwidth]{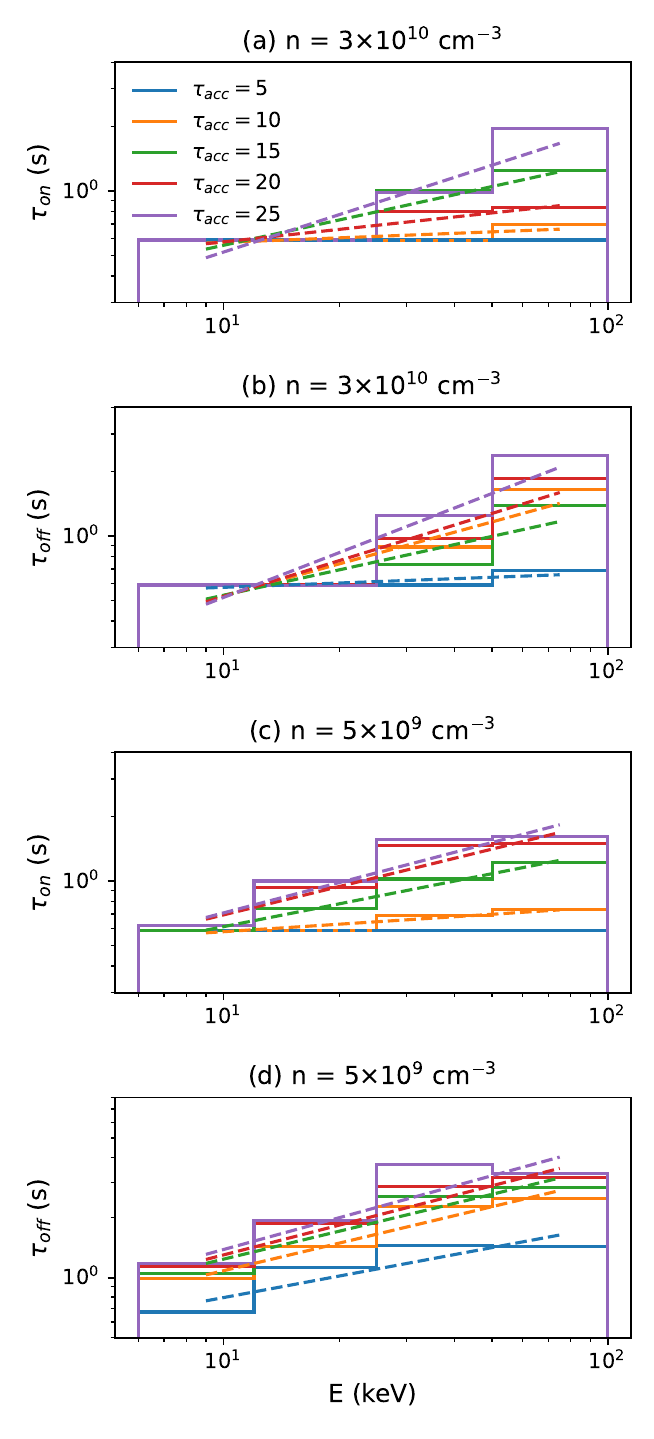}
	\caption{Panels (a) and (c) show the response time scale, $\tau_{\rm{on}}$ and panels (b) and (d) the response time scale, $\tau_{\rm{off}}$, measured at the loop top for cases 5 - 14 (Table \ref{tab:setup}). Continuous lines represent the response time scale measured from Fig.\ \ref{fig:energy-band-hat-tau5}(c) and (d). Dashed lines represent exponential fittings described in Eq.\ (\ref{eq:fitting}) and Fig.\ \ref{fig:fitting} in Appendix \ref{appendix-fitting}.}
	\label{fig:hat-fitting}
\end{figure}
Figure \ref{fig:fitting} in Appendix \ref{appendix-fitting} details how the fitting was done for each simulation case. 
We can see that the sigmoid function shows a good fit with $nVF$ at the onset of acceleration for cases 5 and 9, for higher density cases. The lower density cases have a small overshoot in $nVF$ when the acceleration turns on, but even with the overshot the fitting displays a good agreement with the $nVF$ signal. When the acceleration ends, we observe a good fit up to approximately $t\approx33$ s for lower density cases. After this point, the decrease in $nVF$ occurs slower than the exponential fitting.

The energy dependence of these timescales is summarized in Figures \ref{fig:hat-fitting}a and \ref{fig:hat-fitting}b for high-density cases in the loop-top regions only. We choose to examine the loop tops because it was easier to isolate the response lag from other effects compared to the chromosphere. 
Figure \ref{fig:hat-fitting}a reveals that $\tau_{\rm{on}}$ increases systematically with energy, following an exponential trend with steeper growth rates for shorter acceleration timescales ($\tau_{\rm{acc}}$). Figure \ref{fig:hat-fitting}b demonstrates that $\tau_{\rm{off}}$ is the same order or exceeds $\tau_{\rm{on}}$ across all energy bands, indicating slower relaxation to equilibrium following acceleration cessation. Both timescales exhibit stronger energy dependence for small $\tau_{\rm{acc}}$ values, reflecting faster electron dynamics under more impulsive acceleration conditions. Figures \ref{fig:hat-fitting}c and \ref{fig:hat-fitting}d contain the same analysis for lower density. The lower density cases also manifest an increase in the response lag with the energy, and $\tau_{\rm{off}}$ is also larger than $\tau_{\rm{on}}$. Figures \ref{fig:hat-fitting}a and \ref{fig:hat-fitting}d also contain an exponential fit shown in dashed lines given by 
\begin{equation}\label{eq:fitting}
fit = \mathcal{A}E^{-\mathcal{D}}
\end{equation}
where $\mathcal{A}$ and $\mathcal{D}$ are fitting coefficients. Table \ref{tab:hat} summarizes the results considering $\tau_{\rm{on}} = fit$ and $\tau_{\rm{off}} = fit$, for different acceleration time scales. It is noticed that $\mathcal{D}$ gradually increases with $\tau_{\rm{acc}}$ at the loop top, indicating that larger $\tau_{\rm{acc}}$ values require more time to reach a new steady state. 
The high density cases present a larger slope variation, $\mathcal{D}$, than observed in lower density.

The observed increase in $\tau_{\rm{off}}$ and $\tau_{\rm{on}}$ with energy (Figure \ref{fig:hat-fitting}) arises because electron acceleration does not occur instantaneously. In order for electrons to transition from a lower energy band to a higher energy band, they must pass through intermediate energy bands. Consequently, this process results in a prolonged acceleration time that is energy-dependent. 
\begin{table}[htb]
\centering
\begin{tabular}{lrcccc} \hline\hline
                              & $\tau_{\rm{acc}}$ & \multicolumn{2}{c}{$n=3\times10^{10}(\text{cm}^{-3})$} & \multicolumn{2}{c}{$n=5\times10^{9}(\text{cm}^{-3})$}  \\
                              &   (s)   & $\mathcal{A}(\text{s/keV})$            & $\mathcal{D}$       & $\mathcal{A}(\text{s/keV}) $   & $\mathcal{D}$\\ \hline
\multirow{5}{*}{\stackunder{$\tau_{\rm{on}}$}{(s)}}& 5 & 0.589 & 0.000 & 0.589 & 0.000 \\
                              & 10 &   0.487 & 0.071 & 0.443 & 0.117 \\
                              & 15 &   0.225 & 0.393 & 0.270 & 0.355 \\
                              & 20 &   0.370 & 0.194 & 0.252 & 0.440 \\
                              & 25 &   0.136 & 0.581 & 0.240 & 0.471 \\ \hline
\multirow{5}{*}{\stackunder{$\tau_{\rm{off}}$}{(s)}} & 5 & 0.492 & 0.067 & 0.348 & 0.358 \\
                              & 10 & 0.168 & 0.494 & 0.377 & 0.459 \\
                              & 15 & 0.213 & 0.394 & 0.425 & 0.465 \\
                              & 20 & 0.145 & 0.555 & 0.415 & 0.496 \\
                              & 25 & 0.104 & 0.696 & 0.409 & 0.529 \\ \hline
\end{tabular}
	\caption{Fitting coefficients from Eq.\ (\ref{eq:fitting}) applied to the loop top. Fitting is applied to $nVF$ for a hat pulse, $g_2$, for cases 5 to 14 in Table \ref{tab:setup}.}
\label{tab:hat}
\end{table}

\subsection{Acceleration by sinusoidal wave}  \label{sec:sine-wave}
\subsubsection{$nVF$ evolution}
\begin{figure*}[p]
	\centering
	\includegraphics[width=0.99\textwidth]{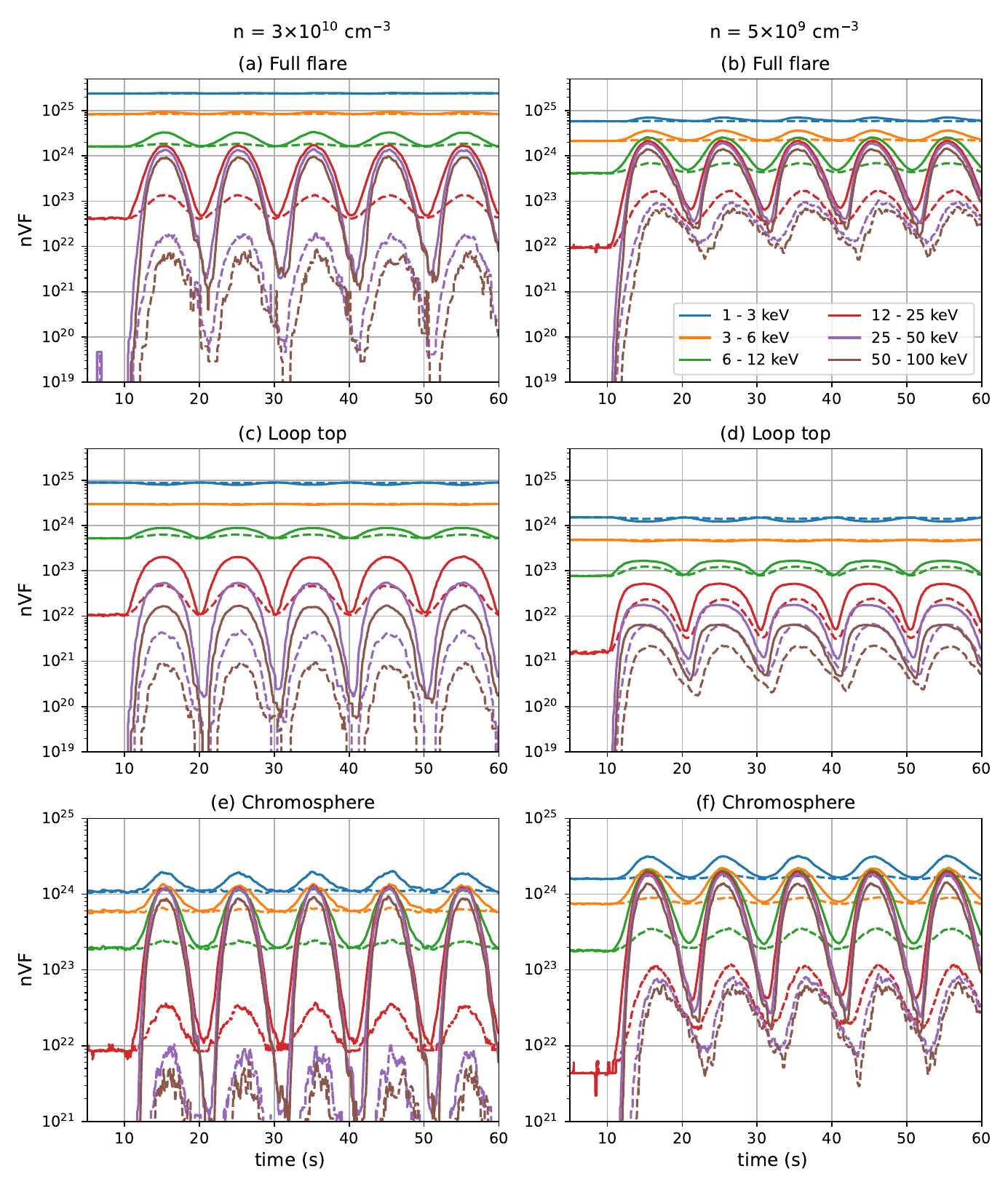}
	\caption{$nVF$ at different energy bands for a sinusoidal  wave (source $g_3$, Eq.\ (\ref{eq:g3})), cases 15-18 (Table \ref{tab:setup})). Continuous lines show results for $\tau_{\rm{acc}}=$ 5 s, dashed lines show results for $\tau_{\rm{acc}}=$ 25 s. Left column, panels (a), (c) and (e), contain results for $n=3\times 10^{10}$ cm$^{-3}$, while right column, panels (b), (d) and (f) contain results for $n=5\times 10^{9}$ cm$^{-3}$.}
	\label{fig:sine-wave-nvf}
\end{figure*}

After analyzing the electron response to a square pulse, this analysis focuses on the electron acceleration and transport properties when subjected to a sinusoidal wave from the source $g_3$ (see Figure \ref{fig:g}).
$nVF$ for cases 15 to 18 in Table \ref{tab:setup} are illustrated in Figure \ref{fig:sine-wave-nvf}. Figures \ref{fig:sine-wave-nvf}a,  \ref{fig:sine-wave-nvf}c and  \ref{fig:sine-wave-nvf}e correspond to the high-density cases, while Figures \ref{fig:sine-wave-nvf}b,  \ref{fig:sine-wave-nvf}d and  \ref{fig:sine-wave-nvf}f contain the low-density cases. The continuous lines represent an acceleration time, $\tau_{\rm{acc}} = 5$ s, and the dashed lines indicate $\tau_{\rm{acc}} = 25$ s.

In both low and high-density cases, the full flare spectra, (Figures \ref{fig:sine-wave-nvf}a and \ref{fig:sine-wave-nvf}b) exhibit a clear signature of $g_3$ in $nVF$. Oscillations are more visible in energy bands above 6 keV for the higher density case and above 3 keV for the lower-density case. The loop top displays a result similar to that seen in the full flare, with one significant difference: energy depletion in the 1-3 keV range, which aligns with results observed for the square pulse. The source signature is also present in the chromosphere (Figures \ref{fig:sine-wave-nvf}e and \ref{fig:sine-wave-nvf}f) where it is more pronounced in the lower energy bands (1-3 keV and 3-6 keV) than the full flare and loop top (Figures \ref{fig:sine-wave-nvf}c and \ref{fig:sine-wave-nvf}d). It is seen that the difference in $nVF$ amplitude between $ \tau_{\rm{acc}} = 5 $ s and $\tau_{\rm{acc}} = 25$ s decreases in the low-density, this is observed in the full flare, loop top and chromosphere. 

In the energy bands of 25-50 keV and 50-100 keV for $ \tau_{\rm{acc}} = 25 $ s, the behavior deviates from that of a sinusoidal wave, possibly due to deceleration in the chromosphere. The oscillation period detected in each energy band, approximately 10 s, coincides with the input signal from $ g_3 $. Additionally, the amplitude of $ nVF $ is considerably larger for $ \tau_{\rm{acc}} = 5 $ s than for $ \tau_{\rm{acc}} = 25 $ s.

In lower-density cases, there is larger lag between the wave signature observed in high and lower energy bands. In lower density cases there is also an increase in the asymmetry between $\tau_{\rm{on}}$ and $\tau_{\rm{off}}$.
The asymmetry in the wave form is seen in Figures \ref{fig:sine-wave-nvf}d and \ref{fig:sine-wave-nvf}f for 25-50 keV and 50-100 keV.
In a real QPP, this lag has been observed in different energy bands, with small variations in period \citep{Zimovets2021}.

\subsubsection{Time lag estimation via cross-correlation} \label{sec:sine-wave-lag}
\begin{figure}[htb]
	\centering
	\includegraphics[ width=0.99\columnwidth]{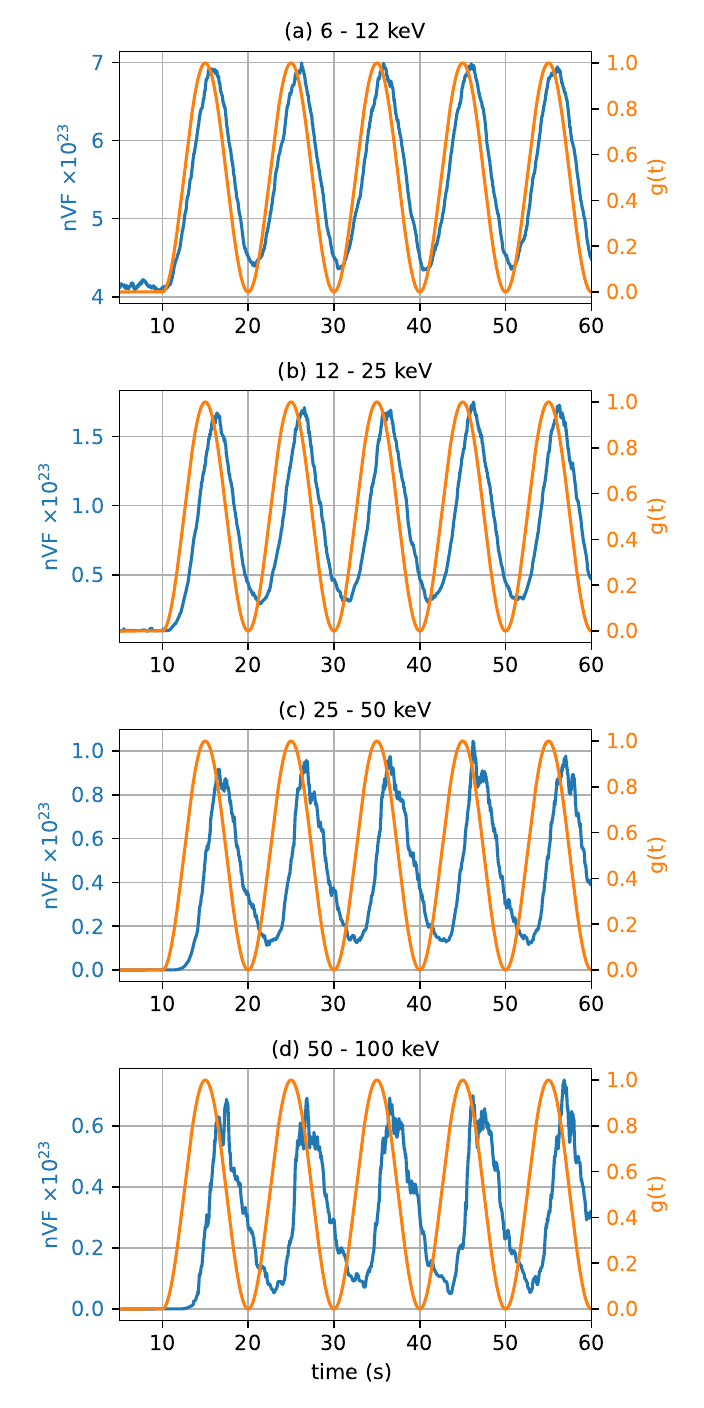}
	\caption{$nVF$ (blue) measured for the full flare and $g_3$ (orange) for simulation case 18 (Table \ref{tab:setup}), where $\tau_{\rm{acc}}=$ 25 s, $n=5\times10^{9}$ cm$^{-3}$.}
	\label{fig:lag-sin}
\end{figure}

Figure \ref{fig:lag-sin} compares the $nVF$ energy band with the input source for a full flare at $\tau_{\rm{acc}} = 25$ s under low density conditions. It is observed that the oscillatory pattern in the 6-12 keV energy band closely resembles the function $g(t)=g_3$. However, as the electrons move to higher energy bands, 25-50 and 50-100 keV, the $nVF$ oscillation pattern begins to diverge from the source oscillation pattern, revealing a lag between $g_3$ and $nVF$.

In order to quantify the acceleration lag across the multiple energy bands, the maximum cross correlation coefficient between $g_3$ and $nVF$ is calculated. Given two discrete-time signals $x[n]$ and $y[n]$ with zero mean, normalized by their variance, their cross-correlation, $R_{xy}[k]$, is computed as
\begin{equation}
R_{xy}[k] = \sum_{n {\rightarrow}   -\infty}^{\infty} x[n]  y[n + k] ,
\end{equation}
where $k$ is the lag (time shift) between the two signals. The summation is taken over all samples where $x[n]$ and $y[n + k]$ overlap. The time lag $\Delta t$ is given
\begin{equation}
\Delta t = \frac{\hat{k}}{f_s} ,
\label{eq:lag}
\end{equation}
where the optimal lag $ \hat{k} $ is the index that maximizes the cross-correlation $R_{xy}[k]$ and $f_s$ is the sampling rate. 

\begin{figure*}[htb]
	\centering
	\includegraphics[width=.99\textwidth]{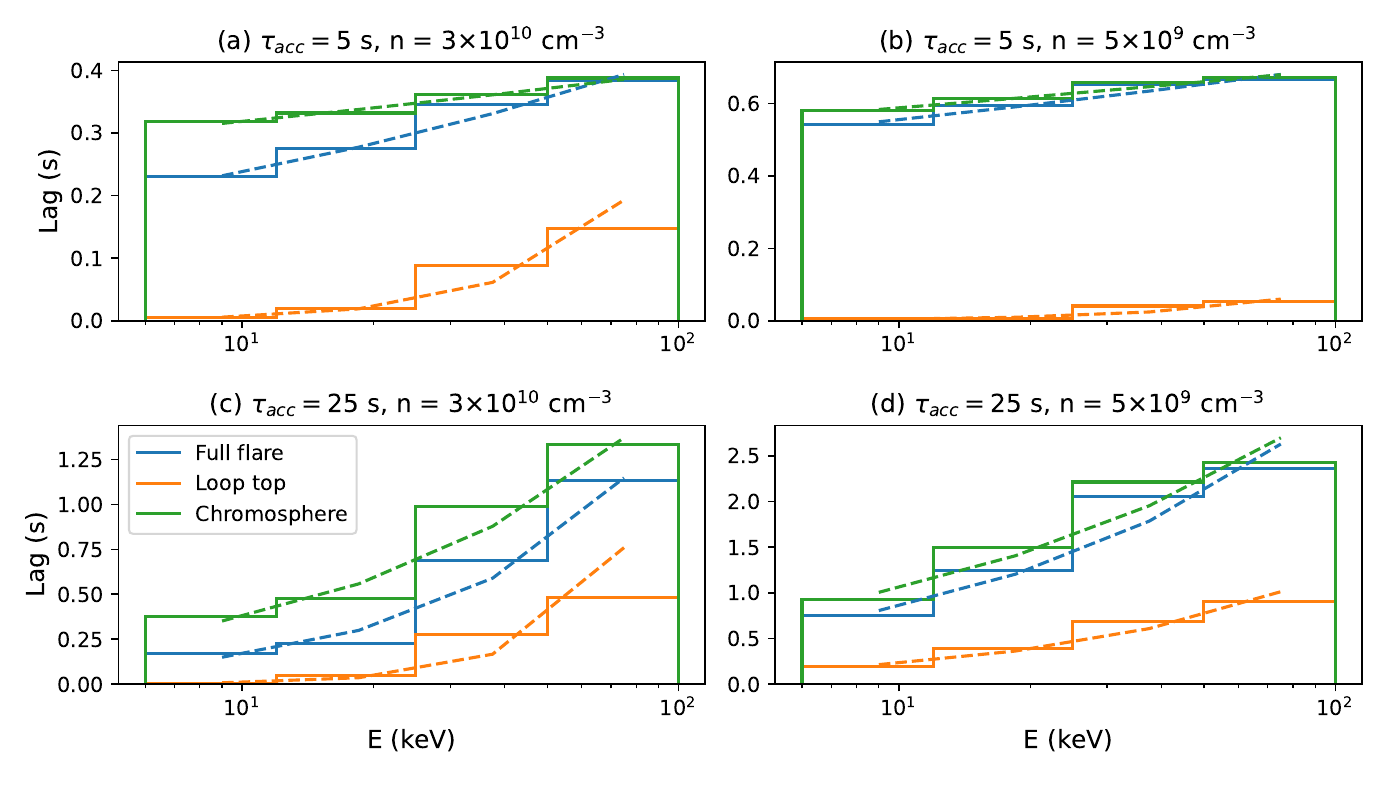}
	\caption{Response time scale Eq.\ (\ref{eq:lag}) computed for a sinusoidal wave (source $g_3$, Eq.\ (\ref{eq:g3})) obtained from the $nVF$ time series shown in Fig.\ \ref{fig:sine-wave-nvf}. Panels (a) and (c) contain results for higher plasma number density and panels (b) and (c) for lower plasma number density. The dashed lines represent a fit using an exponential function, Eq.\, (\ref{eq:fitting}). }
	\label{fig:sin-lag}
\end{figure*}

Figure \ref{fig:sin-lag} shows the variation of the time lag for source $g_3$ across different energy bands, loop regions and $\tau_{\rm{acc}}$. Firstly, it is observed that the time lag increases exponentially with higher energy bands. This trend is consistent across all turbulent acceleration time scales and plasma number densities. At the loop top, the maximum lag occurs for cases with $ \tau_{\rm{acc}} = 25 $ s, reaching 0.9 s for the lower density case and 0.5 s for the higher density case. In contrast, the lag is negligible for $ \tau_{\rm{acc}} = 5 $ s, remaining below 0.2 s. 

When the electrons move to the chromosphere, the lag increases to 2.4 s and 1.3 s for $ \tau_{\rm{acc}} = 25 $ s at low and high densities, respectively. This highlights that the chromospheric time lag dominates the lag observed in overall flare signatures. For $\tau_{\rm{acc}} = 5$ s in the chromosphere, the lag is shorter, 0.4 s for high density 0.6 s for low density, making it negligible for most instruments. Table \ref{tab:wave-sin} summarizes the coefficients for the exponential fit (Eq.\ (\ref{eq:fitting})) used to estimate the time lag, as shown in Fig. \ref{fig:sin-lag}.
\begin{table}[htb]
\centering
\begin{tabular}{lcccc}  \hline\hline
\multirow{2}{*}{$\tau_{\rm{acc}}$ (s)} & \multirow{2}{*}{n  (cm$^{-3}$) }   & \multicolumn{3}{c}{$\mathcal{D}$}         \\
                        &            & Full flare & Loop top & Chromosphere \\ \hline
5 &  3$\times10^{10}$  & 0.2498     & 1.6516   & 0.0949       \\
25 &  3$\times10^{10}$ & 0.9612     & 2.1895   & 0.6426       \\
5 &  5$\times10^{9}$   & 0.1007     & 1.2990   & 0.0717       \\
25 &  5$\times10^{9}$  & 0.5579     & 0.7341   & 0.4657       \\ \hline
\end{tabular}
\caption{Coefficients obtained for the exponential fit, Eq.\ (\ref{eq:fitting}), applied to the time lag shown in Fig. \ref{fig:sin-lag}.}
\label{tab:wave-sin}
\end{table}

\subsubsection{Spectral index evolution}
\begin{figure*}[p]
	\centering
	\includegraphics[width=0.99\textwidth]{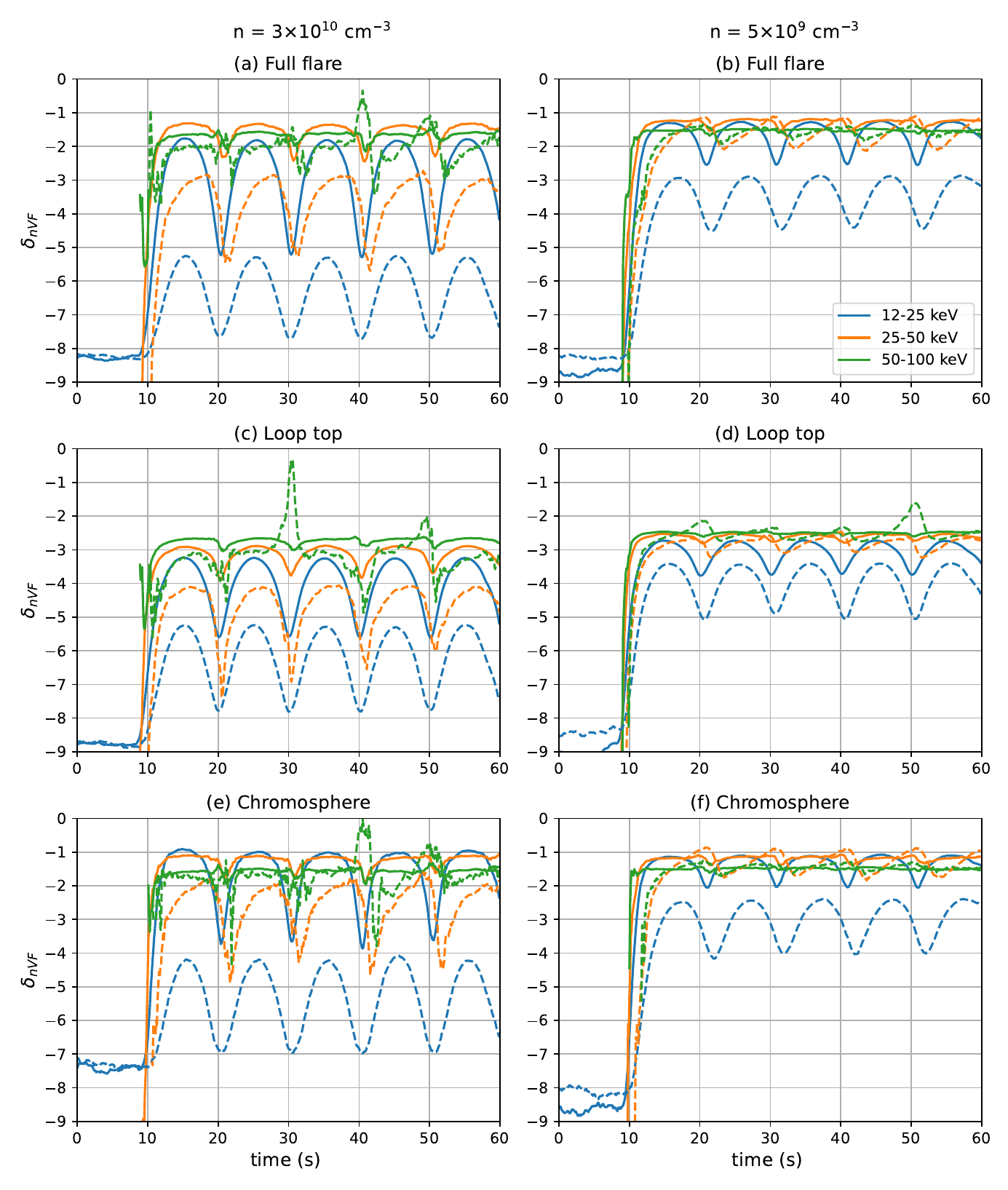}
	\caption{Spectral index at different energy bands for a sinusoidal  wave (source $g_3$, Eq.\ (\ref{eq:g3})) for cases 15-18 (Table \ref{tab:setup}). Continuous lines show results for $\tau_{\rm{acc}}=$ 5 s, dashed lines show results for $\tau_{\rm{acc}}=$ 25 s. Left column panels (a), (c) and (e), contain results for $n=3\times 10^{10}$ cm$^{-3}$, while right column panels (b), (d) and (f) contain results for $n=5\times 10^{9}$ cm$^{-3}$. $\delta_{nVF}$ was computed using data from Fig.\ \ref{fig:sine-wave-nvf}.}
	\label{fig:spectral-index-sine}
\end{figure*}

Figure \ref{fig:spectral-index-sine} presents the spectral index variation integrated over 4 seconds for cases 15 and 16 in Figures \ref{fig:spectral-index-sine}a, \ref{fig:spectral-index-sine}c and \ref{fig:spectral-index-sine}e, as well as Figures \ref{fig:spectral-index-sine}b, \ref{fig:spectral-index-sine}d and \ref{fig:spectral-index-sine}f for cases 17 and 18. In the plots, the solid line represents $\tau = 5$ s, while the dashed lines indicate $\tau = 25$ s.

In the left panels (Figures \ref{fig:spectral-index-sine}a, \ref{fig:spectral-index-sine}c and \ref{fig:spectral-index-sine}e) representing the high-density case, a sinusoidal wave signature is observed in the 12-25 keV range for $\tau = 25$ s along the multiple regions of the flare. As electrons move to higher energy bands, the wave shape gradually transitions to a flatter pattern with smaller amplitude variations. At the chromosphere and full flare, the 25-50 keV wave pattern resembles a sawtooth shape.

It is important to note that the acceleration begins at $t = 10$ s, and the oscillation period is 10 s. Even when the turbulent acceleration decreases to zero at $t = 20$, 30, 40, and 50 s, the spectral index does not return to its original steady state. Thus, energy spectra here do not return to a thermal distribution. This is consistent with the relaxation time scale $\tau_{\rm{off}}$, discussed for case $g_2$. It takes longer for the accelerated electrons to lose their energy and exit the loop entirely.

Figures \ref{fig:spectral-index-sine}b, \ref{fig:spectral-index-sine}d and \ref{fig:spectral-index-sine}f display results for the low-density cases, where it is seen that the oscillation amplitudes are smaller during the full flare, loop top and chromosphere than those observed in the high-density cases. This observation aligns with the trends noted in the temporal evolution of the $nVF$, as shown in Fig. \ref{fig:sine-wave-nvf}. Note that all behavior discussed in the high-density cases are observed in the low-density cases. As an example, 
in the range of 12-25 keV for the low density case the spectral index oscillates from -4.5 to -3 for the full flare spectra (Figure \ref{fig:spectral-index-sine}b). This is similar to a `soft-hard-soft' behavior often observed in solar flares (e.g., \cite{kane_spectral_1970,grigis_spectral_2008}) where the spectral index is steeper, flatter and the steeper again, which is a signature of the acceleration process as discussed by \cite{kane_spectral_1970}. \LAS{While soft-hard-soft behavior is observed in flare emissions, it is not periodic. In the context here, it is periodic due to the nature of our driver, which has a specific periodic characteristic, and the loop model that maintains constant properties over time, such as temperature, number density and length. However, these properties can change significantly during a flare event, and we expect this to therefore change the resultant `soft-hard-soft' behavior \citep{fletcher_observational_2011}.}


\subsection{Response to a damped wave} \label{sec:damped-wave}
\subsubsection{$nVF$ evolution}

Our final transient case examines a damped wave (source $g_4$ described in Eq.\ (\ref{eq:g4})) which consists of a sinusoidal wave with exponential decay. Source $g_4$ represents electron acceleration either by a wave packet reaching the loop top \citep{pascoe_spatial_2012} or by an oscillatory damped driver, such as oscillatory reconnection, e.g. \citep{McLaughlin2009, schiavo2024energymap,schiavo2024PoP}. The pulse begins at $t = 10$ s, with damping starting at $t = 15$ s.

\begin{figure*}[p]
	\centering
	\includegraphics[width=0.99\textwidth]{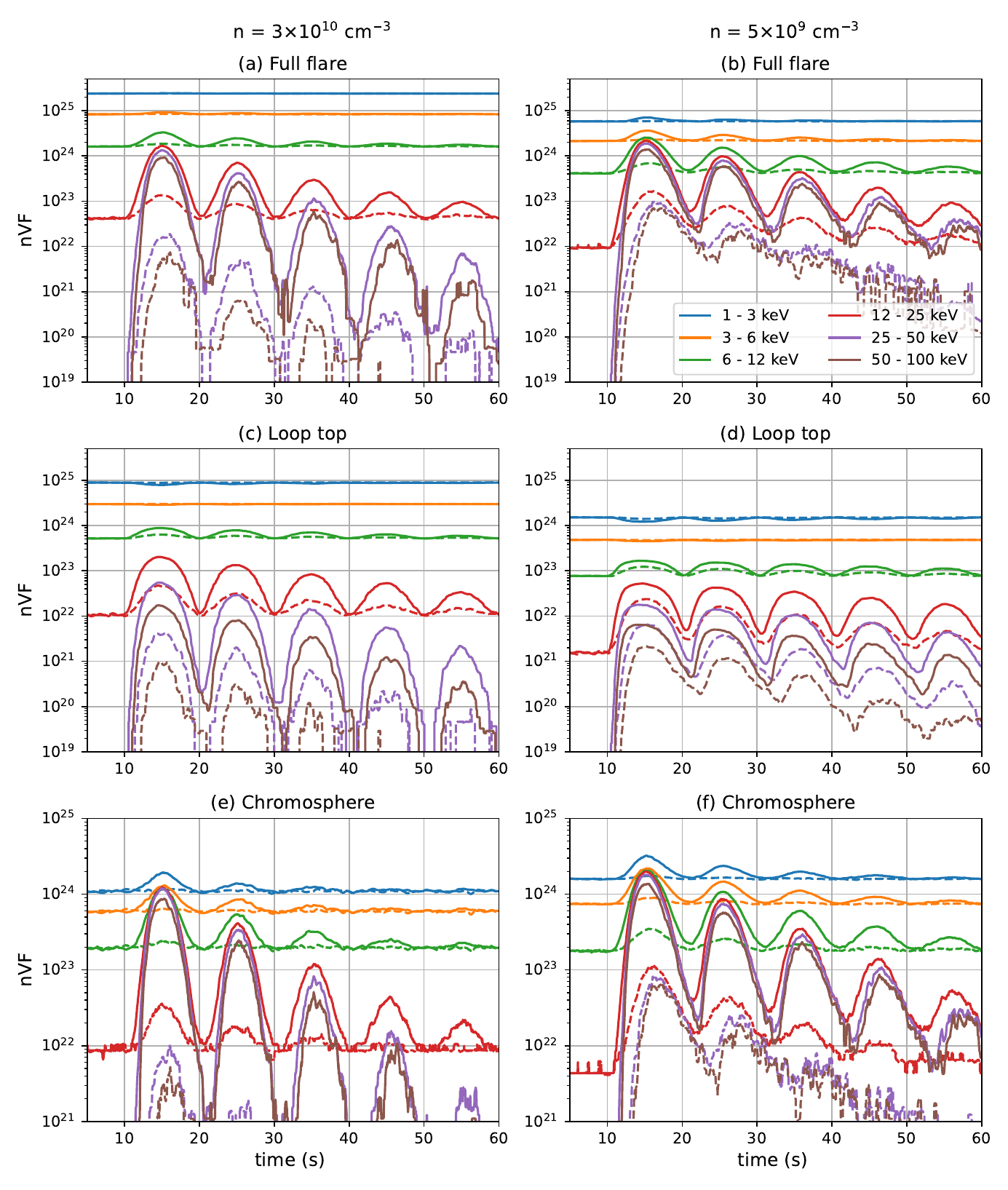}
	\caption{$nVF$ at different energy bands for a damped wave (source $g_4$, Eq.\ (\ref{eq:g4})). Continuous lines show results for $\tau_{\rm{acc}}=$ 5 s, dashed lines show results for $\tau_{\rm{acc}}=$ 25 s. Left column panels (a), (c) and (e), contain results for $n=3\times 10^{10}$ cm$^{-3}$, while right column panels (b), (d) and (f) contain results for $n=5\times 10^{9}$ cm$^{-3}$.}
	\label{fig:damped-sine-wave-nvf}
\end{figure*}

Figure \ref{fig:damped-sine-wave-nvf} presents the $nVF$ curves integrated over 4 s for cases 19-22 in Table \ref{tab:setup}. In the figure, solid lines represent $\tau_{\rm{acc}} = 5$ s, while dashed lines indicate $\tau_{\rm{acc}} = 25$ s. In Figures \ref{fig:damped-sine-wave-nvf}a, \ref{fig:damped-sine-wave-nvf}c and \ref{fig:damped-sine-wave-nvf}e, which correspond to higher density cases, several key features emerge. Firstly, clear response to the turbulent acceleration appear in $nVF$ at energies above 6 keV in both the full flare (\ref{fig:damped-sine-wave-nvf}a) and loop top (\ref{fig:damped-sine-wave-nvf}c) regions. Secondly, the chromospheric oscillations in Figure \ref{fig:damped-sine-wave-nvf}e accurately reproduce the driver period across all energy bands (1 to 100 keV) even when the driver amplitude is small. Thirdly, the $nVF$ curves exhibit exponential decay consistent across all energy bands seen in Figures \ref{fig:damped-sine-wave-nvf}a to \ref{fig:damped-sine-wave-nvf}f. Finally, the amplitude of the oscillations shows a significant dependence on $\tau_{\rm{acc}}$, with $\tau_{\rm{acc}}=5$ s amplitudes larger than those for $\tau_{\rm{acc}}=25$ s. This is consistent with behavior observed in previous cases.

The right panels (Figures \ref{fig:damped-sine-wave-nvf}b, \ref{fig:damped-sine-wave-nvf}d and \ref{fig:damped-sine-wave-nvf}f) for lower-density cases, display similar behavior to the high density cases except at the full flare and chromosphere, where oscillations after $t=35$ s disappear in the 25-50 keV and 50-100 keV energy bands.

\subsubsection{Time lag estimation for a damped wave}
\begin{figure}[htb]
	\centering
	\includegraphics[ width=0.99\columnwidth]{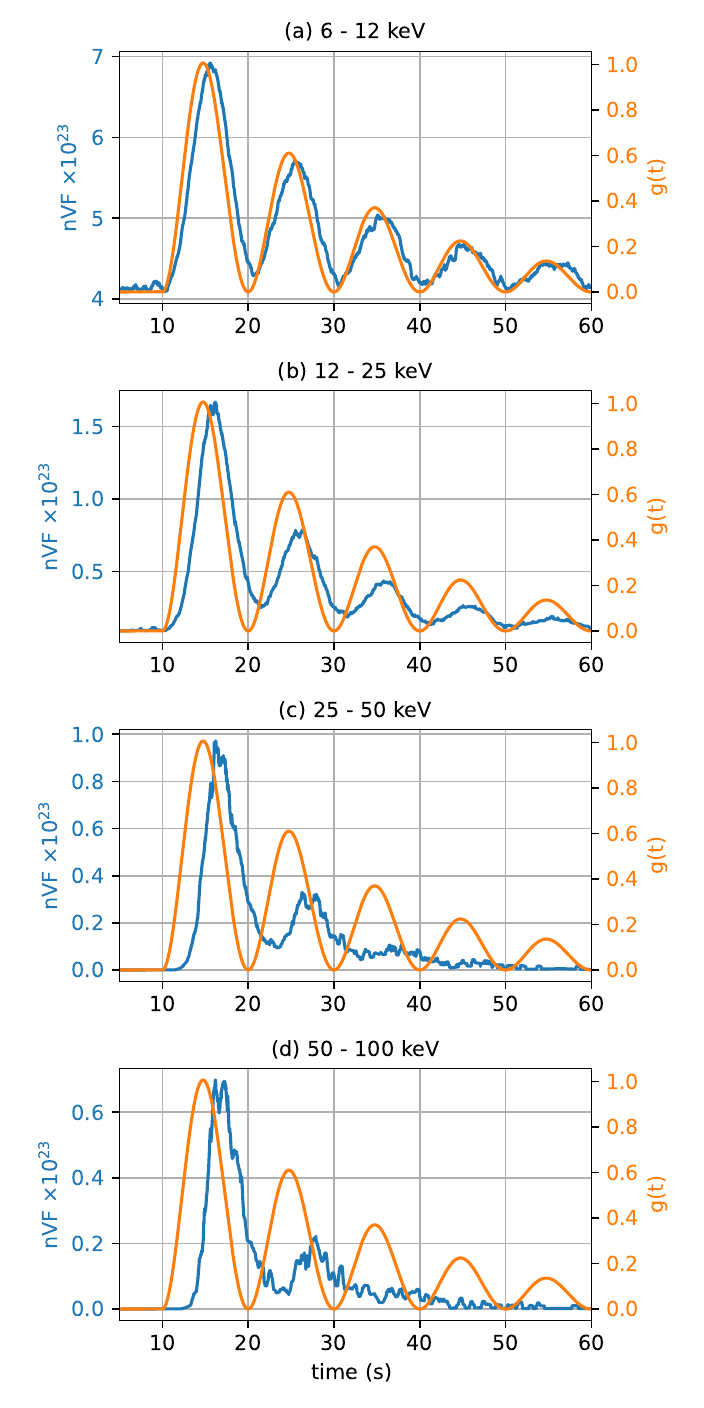}
\caption{$nVF$ (blue) measured for the full flare and $g_4$ (orange) for simulation case 22, where $\tau_{\rm{acc}}=$ 25 s, $n=5\times10^{9}$ cm$^{-3}$.}.
	\label{fig:lag-damping}
\end{figure}

Figure \ref{fig:lag-damping} compares the $nVF$ energy bands with the input source function $g_4$ for a full flare under low-density conditions, with an acceleration timescale of $\tau_{\rm{acc}} = 25$ s for the damped wave pulse. The 6-12 keV energy band closely follows the oscillatory pattern of $g_4$, maintaining excellent synchronization in both phase and amplitude. However, in the higher energy bands (25-50 keV and 50-100 keV), two distinct deviations are observed, a noticeable phase lag emerges between $g_4$ and the $nVF$ response, and the wave pattern disappears entirely after $t=35$ s, when the $g_4$ amplitude is smaller than 35\% of the original amplitude. The same effect is not seen in lower energy bands 6-12 keV or 12-25 keV. This means that soft X-rays and lower energy hard X-rays are more sensitive to the turbulent acceleration. This is seen even in small perturbation amplitudes less than $0.2$.

In Figure \ref{fig:lag-damping}, the $nVF$ response lag is quantified for source $g_4$ across different loop regions, similar to the analysis for the sinusoidal  wave in Section \ref{sec:sine-wave-lag}. It was  observed that the time lag exhibits exponential growth as energy increases, resembling the behavior seen in non-damped wave cases in Section \ref{sec:sine-wave-lag}. This exponential relationship remains consistent across all tested turbulent acceleration timescales and plasma densities. The chromospheric region consistently shows larger time lags than other loop regions. 
The exponential fit function, Eq.\ (\ref{eq:fitting}), was applied to the response timescale and the trends are summarized in Table \ref{tab:wave-damped}. This function is applied in modeling the growth of $nVF$ response lag as function of energy which is shown in Figure \ref{fig:lag-damping}. These coefficient $\mathcal{D}$ characterizes how the time lag scales with energy under various physical conditions, providing information on how electrons are accelerate across different energy bands.
\begin{table}[htb]
\centering
\begin{tabular}{lcccc}  \hline\hline
\multirow{2}{*}{$\tau_{\rm{acc}}$ (s)}  & \multirow{2}{*}{n  (cm$^{-3}$)}   & \multicolumn{3}{c}{$\mathcal{D}$}         \\
                        &            & Full flare & Loop top & Chromosphere \\ \hline
5 &  3$\times10^{10}$  & 1.3741     & 0.0000   & 0.3449       \\
25 &  3$\times10^{10}$ & 2.8412     & 2.0106   & 0.8139       \\
5 &  5$\times10^{9}$   & 0.2278     & 1.8168   & 0.1727       \\
25 &  5$\times10^{9}$  & 0.7582     & 1.2693   & 0.6221      \\ \hline
\end{tabular}
\caption{Coefficients obtained for the exponential fit applied to the response lag shown in Fig.\ \ref{fig:lag-damping}, using Eq.\ (\ref{eq:fitting}). }
\label{tab:wave-damped}
\end{table}


\subsubsection{Damping rate}
To quantify the damping rates observed in $nVF$, the $nVF$ values are normalized to a range between 0 and 1. After normalizing an exponential function is fitted to the peaks of the $nVF$, extracting the damping envelope described by $\exp[-(t-15)/\tau_{\rm{exp}}]$, where $\tau_{\rm{exp}}$ represents the exponential decay timescale.

\begin{figure*}[htb]
	\centering
\includegraphics[width=0.98\textwidth]{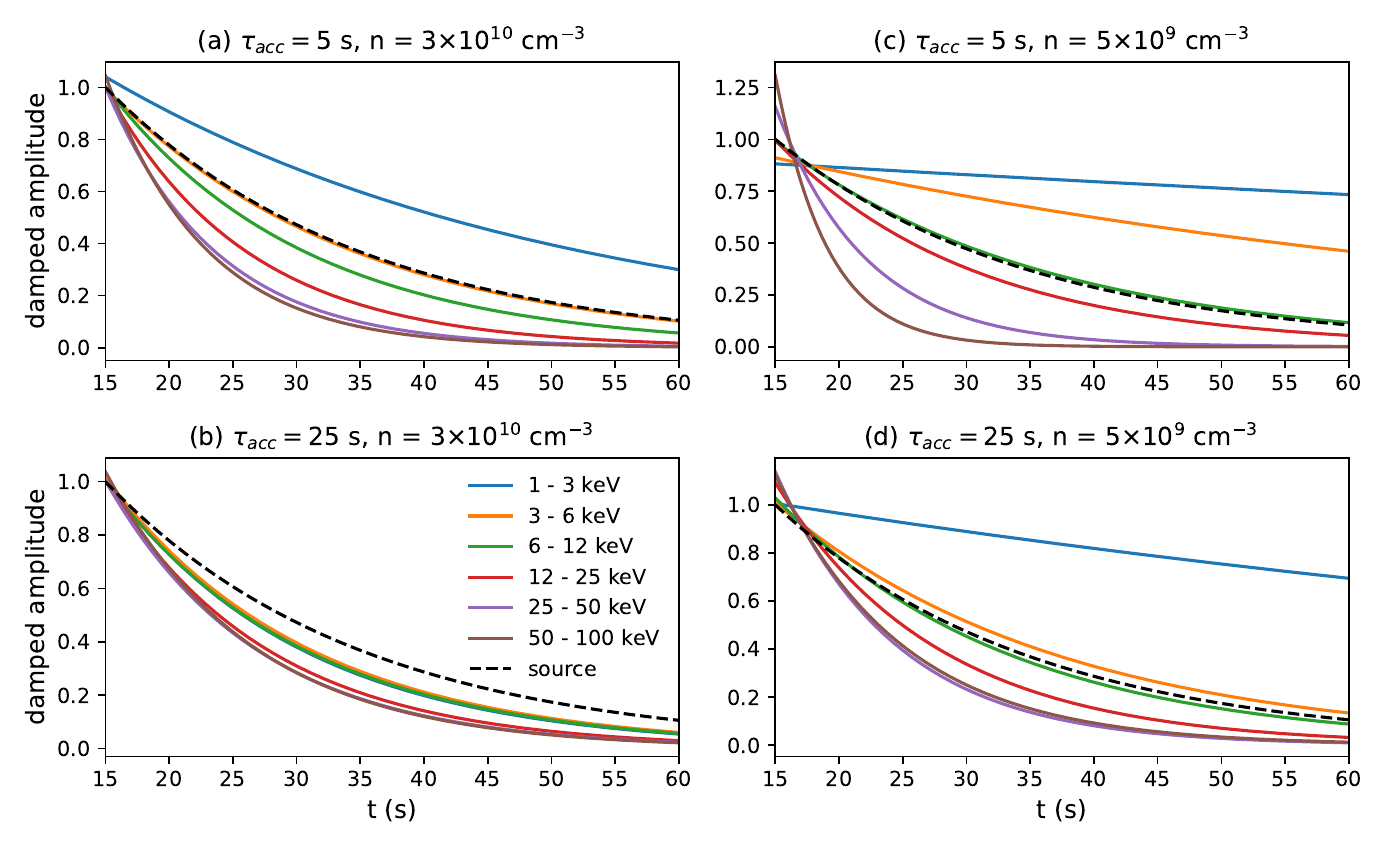}
	\caption{Normalized damping envelope, $\exp[-(t-15)/\tau_{\rm{exp}}]$, for $nVF$ time series extracted from Fig.\ \ref{fig:damped-sine-wave-nvf}, for simulation cases 19 to 22 in Table \ref{tab:setup}.}
	\label{fig:fitting-damping}
\end{figure*}

Figure \ref{fig:fitting-damping} displays the fitted damping envelopes for simulations 19 through 22 in Table \ref{tab:setup}. These results reveal important trends and diagnostics. Firstly, the damping rate systematically increases with higher energy bands, seen in Figures \ref{fig:fitting-damping}a to \ref{fig:fitting-damping}d, where the fastest decay occurs at 50-100 keV. Secondly, comparing Figures \ref{fig:fitting-damping}c and \ref{fig:fitting-damping}d at 25-50 and 50-100 keV it is seen that shorter turbulent acceleration timescales exhibit faster decay. Finally, reducing the plasma number density significantly enhances the damping rate, seen in Figures \ref{fig:fitting-damping}a and \ref{fig:fitting-damping}c at 50-100 keV.

\begin{figure}[htb]
	\centering
	\includegraphics[width=0.99\columnwidth]{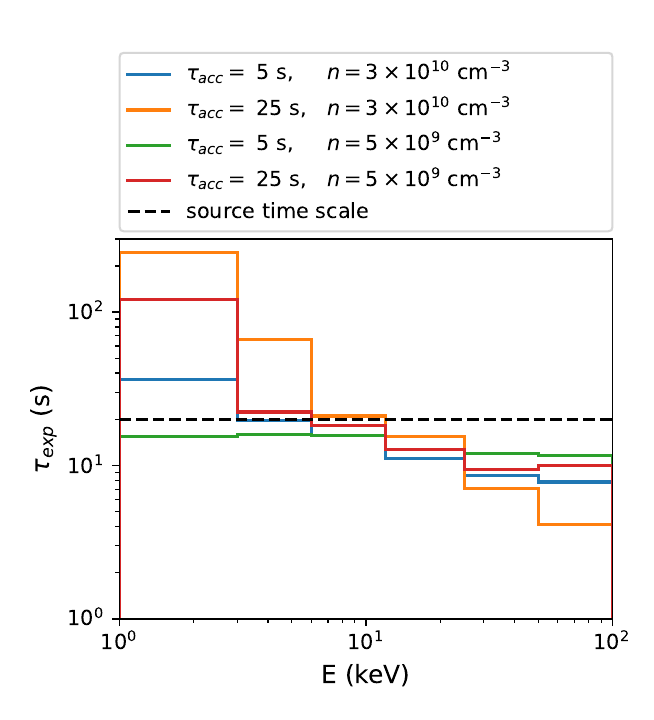}
	\caption{Exponential damping time scale extracted from Fig.\ \ref{fig:fitting-damping}, for cases 19 to 22, Table \ref{tab:setup}.}
	\label{fig:fitting-damping-summary}
\end{figure}

The summary plot in Figure \ref{fig:fitting-damping-summary} illustrates these relationships across the entire flare region. Lower energy bands show slower decay rates, as larger the $\tau_{\rm{exp}}$ values indicate. 

Furthermore, the variation of $\tau_{\rm{exp}}$ across different energy bands becomes more pronounced (with a steeper gradient) when $\tau_{\rm{exp}}$ increases. The same is seen when plasma number density increases the variation of $\tau_{\rm{exp}}$ with energy is also enhanced. Although both the increasing of $\tau_{\rm{acc}}$ and plasma number density enhances the $\tau_{\rm{exp}}$ variation, the system is more sensitive to variations on $\tau_{\rm{acc}}$.

\section{Conclusions}\label{sec:conclusions}

This study investigated electron acceleration and transport in a flaring coronal loop by numerically solving a time-dependent Fokker-Planck equation. Our model incorporates transient acceleration mechanisms by adding an amplitude modulation in the turbulent diffusion coefficient (Eq.\ \ref{eq:Dturb}). It simulates an unsteady energy input, such as from a pulse or wave train, to emulate the dynamics of transient reconnection processes (e.g., oscillatory reconnection) at the loop apex. We characterized the density-weighted electron flux, a diagnostic directly comparable to observed X-ray emissions, across both energy and spatial domains, from the corona to the chromosphere. We modeled a 40$^{\prime\prime}$ symmetric flaring coronal loop, where the loop apex was placed at $z=0^{\prime\prime}$ and the chromosphere at $|z|=20^{\prime\prime}$, the electrons were accelerated and replenished at $|z|<3^{\prime\prime}$.

A key focus of our analysis was the electron response timescale when accelerated by a source. 
A detailed parameter study examined how the temporal profile of the turbulent diffusion coefficient, $g(t)$, influences electron acceleration and  transport. Our findings indicate that the density-weighted electron flux is highly sensitive to the pulse shape used to accelerate electrons at the loop top. Additionally, it was determined that the response timescale for a flaring loop at 20 MK is less than 0.6 s at 6-12 keV energy band and less than 3 seconds at 50-100 keV bands. Furthermore, it was demonstrated that the electron response delay is energy dependent, gradually increasing for higher-energy electrons.

To prevent electron depletion and enable a more comprehensive analysis of electron dynamics, an electron replenishment condition was introduced that acts as a continuous supply of electrons at the loop top. This approach provides a robust framework for studying long-term particle behavior and its implications for solar atmospheric heating and particle acceleration processes. Our analysis of replenishment effects shows that the spatial distribution of the injected electrons does not influence the energy spectra and pitch angle distributions, but has a minor influence on electron distribution  on the loop apex $|z|<3^{\prime\prime}$. 

Examining square-pulse acceleration revealed an asymmetry in relaxation timescales. The time required for electrons to adapt to the onset and cessation of acceleration are different, while the relaxation timescale is longer when turbulent acceleration is terminated compared to when it is initiated. Moreover, this relaxation timescale exhibits an exponential increase with electron energy as seen Figure \ref{fig:hat-fitting}.

The sinusoidal and damped sinusoidal sources exhibited a lag between $nVF$ and the source. This response lag was also energy dependent, consistent with observations for the square pulse source. However, for the sinusoidal and damped sinusoidal sources, the $nVF$ was not significantly altered because the response lag was much shorter than the period of the wave source.

It was observed that the temporal signature of the input acceleration profile, $g(t)$, persists in the $nVF$ time signal across all simulations and energy bands. This signature remains detected regardless of plasma number density. However, it becomes less pronounced under low-density conditions at higher energies (25-50 keV and 50-100 keV) and high turbulent-acceleration time scales. This was observed in Figures \ref{fig:energy-band-hat-tau5}, \ref{fig:sine-wave-nvf} and \ref{fig:damped-sine-wave-nvf}.


Analysis of the spectral index evolution in time, Figure \ref{fig:spectral-index-sine}, for the sinusoidal and damped sinusoidal waves revealed that while the transient $nVF$ signal closely follows the input pulse $g(t)$, the spectral index develops a distinct oscillation pattern similar to the `soft-hard-soft' pattern observed in flares, except in the low-energy band (6-12 keV). 
For the sinusoidal and damped sinusoidal cases 
(Figures \ref{fig:sine-wave-nvf} and \ref{fig:damped-sine-wave-nvf}), the spectral index oscillations retained the same frequency as the input signal. The shape of the transient spectral index is highly affected by $\tau_{\rm{acc}}$ and plasma number density. This is caused by the dependency of the response time scale to electron energy and plasma number density.

In the damped wave case, we found good synchronization between the source and $nVF$ (Figure \ref{fig:lag-damping}). However, the wave pattern disappears entirely after $t = 35$ s when the $g_4$ amplitude drops below 35\% in the highest energy bands. This result indicates that electrons stop responding to the acceleration in the higher energy bands when driving amplitude is low.

In the damped wave case, it was found that the exponential decay timescale decreases with energy, shown in Figure \ref{fig:fitting-damping-summary}. Additionally, the energy dependency of the exponential damping time scale ($\tau_{\rm{exp}}$) is influenced by plasma number density and turbulent acceleration time scale ($\tau_{\rm{acc}}$). Enhancing plasma number density and $\tau_{\rm{acc}}$ increase the energy ($E$) dependency, $\partial \: \tau_{\rm{exp}}/\partial E$.

The results of this paper have diagnostic potential in flare observations. 
The shape of the acceleration temporal profile, $g(t)$, manifests across the $nVF$ energy bands and spectral indices as a fingerprint of the source term. 
The electron response time scale increases with the electron energy and electrons stop responding to the acceleration in the higher energy bands when driving amplitude is low.


\LAS{While these studies reveal the details of how turbulence is generated and evolves, its net effect on electron populations is to provide a stochastic acceleration term \citep{chen_determination_2013}. The precise temporal form of this acceleration is complex, potentially involving linear growth and saturation phases \citep{crawford_2024} or transitions between different acceleration regimes \citep{crawford_2025}. Rather than modeling a specific instability, our work here has taken a foundational approach, i.e., simple forms of $g(t)$, and we have investigated how the electron transport system itself responds to a generalized transient driver amplitude. This has  allowed us to isolate and characterize the intrinsic energy-dependent response timescales that will imprint on observations, regardless of the specific underlying turbulent mechanism.}

\LAS{
Future directions for this work include investigating more advanced forms for $g(t)$, such as a two-stage turbulent process: an initially linear exponential growth phase and a later saturated state, e.g., \citet{che_2020}. It should also be noted that the nature of turbulent acceleration is highly sensitive to the local plasma conditions; for instance, the presence of a strong guide field during reconnection can transition the mechanism from a vortical, stochastic acceleration to a more efficient Alfvénic-like Fermi process, producing harder energy spectra \citep{crawford_2025}. These effects could be incorporated into the model by modifying Equation (\ref{eq:Dturb}).
}

In summary, this work presents a novel methodology for analyzing electron acceleration and transport in flares driven by unsteady sources. Our results provide new insights into accelerated electron timescales and energy-dependent behavior, with implications for interpreting solar flare observations and understanding particle transport in astrophysical plasmas. Transport effects such as short timescale scattering were not included in this study but will likely affect the acceleration and decay timescales and this will be tested in future modeling.

\vspace{0.5cm}
All authors acknowledge the UK Research and Innovation (UKRI) Science and Technology Facilities Council (STFC) for support from grant ST/X001008/1 and for IDL support. The research was sponsored by the DynaSun project and has thus received funding under the Horizon Europe programme of the European Union under grant agreement (no. 101131534). Views and opinions expressed are however those of the author(s) only and do not necessarily reflect those of the European Union and therefore the European Union cannot be held responsible for them. This work was also supported by the Engineering and Physical Sciences Research Council (EP/Y037464/1) under the Horizon Europe Guarantee.

\software{NumPy \citep{numpy},  
          SciPy \citep{SciPy}, 
          matplotlib \citep{matplotlib},
          WALSA tools \citep{jafarzadeh_wave_2025}.
          }

%






\appendix
\section{Fokker-Planck equation coefficients} \label{appendix-eq}

Eq.\ (\ref{eq:fp-original}) is written as a function of the electron flux spectra $F(z,E,\mu,t)$ by multiplying it by $v^2/m_e$ and substituting $f=m_eF/v^2$
\begin{eqnarray}
	\pdv{F}{t} =
	-\mu v \pdv{F}{z} + 
	\underbrace{\pdv{}{v}\left[v^2D_{\rm{turb}}\pdv{F/v^2}{v}\right]}_{I}
    &+&  \underbrace{\frac{\Gamma}{2} 
		\pdv{}{v} \left( 2v G(u) \pdv{}{v}\left(\frac{F}{v^2}\right) +
		4u^2 G(u)\frac{F}{v^2}
		\right)}_{II}  \nonumber  \\ 
&+&\underbrace{\frac{\Gamma }{2}  \frac{1}{v^3} \pdv{}{\mu} \left( (1 - \mu^2) \left( \text{erf}(u) - G(u) \right) \pdv{}{\mu}F \right)}_{III} \label{eq:fp-original-g} . 
	\label{eq:fp-original-g}
\end{eqnarray}
A Fokker-Planck equation with $n+1$ variable ($t,x_1,x_2,...,x_n$) can be written as
\begin{equation}
    \pdv{P}{t} = -\frac{\partial}{\partial x_i} (A_i P) + \frac{1}{2} \frac{\partial^2}{\partial x_i x_j}(B_{i,j} P) ,
    \label{spde}
\end{equation}
where $P$ is a distribution function, $A_i$ and $B_{ij}$ are the coefficients of the original partial differential equation, and $\mathbf{x} =(z,v,\mu)$. From Eq.\ (\ref{spde}) we notice that the Fokker-Planck equation can be written as
\begin{eqnarray}
    \pdv{F}{t} =- \frac{\partial}{\partial z} (A_z F) &-&\frac{\partial}{\partial v}(A_v F) + \frac{1}{2}\frac{\partial^2}{\partial v^2}(B_{v} F) -\frac{\partial}{\partial \mu}(A_\mu F) + \frac{1}{2}\frac{\partial^2}{\partial \mu^2}(B_{\mu} F) ,
\label{eq:FP-2}
\end{eqnarray}
where $A_z=\mu v$. Term $I$ from Eq.\ (\ref{eq:fp-original-g}) can be expanded as
\begin{eqnarray}
I &=& \pdv{}{v}\left[v^2D_{\rm{turb}}\pdv{F/v^2}{v}\right] = \pdv{^2}{v^2}\left(D_{\rm{turb}}F\right) -\pdv{}{v}\left[\left(\frac{2D_{\rm{turb}}}{v}+\pdv{D_{\rm{turb}}}{v} \right)F\right]\nonumber  \\
 &=& \frac{1}{2}\pdv{^2}{v^2}\left(\underbrace{2D_{\rm{turb}}}_{B_{\rm{turb}}}F\right) - \pdv{}{v}\left[\underbrace{\left(\frac{2D_{\rm{turb}}}{v}+\pdv{D_{\rm{turb}}}{v} \right)}_{A_{\rm{turb}}}F\right] ,
 \label{eq:term-I}
\end{eqnarray}
Expanding term II from Eq.\ (\ref{eq:fp-original-g})
\begin{eqnarray}
II &=&\frac{\Gamma}{2} 
\pdv{}{v} \left( 2v G(u) \pdv{}{v}\left(\frac{F}{v^2}\right) +
4u^2 G(u)\frac{F}{v^2}
\right) \nonumber \\
&=& \pdv{^2}{v^2} \left( \frac{\Gamma  G(u)}{v} F \right) 
+ \pdv{}{v} \left(2 \Gamma u^2 G(u)\frac{F}{v^2}
-F\pdv{}{v}\left( \frac{\Gamma  G(u) }{v} \right)    
-\frac{2\Gamma G(u)}{v^2}F\right) \nonumber \\
&=& \frac{1}{2}\pdv{^2}{v^2} \left( \frac{2\Gamma  G(u)}{v} F \right) - \pdv{}{v} \left[-\left(
\frac{2 \Gamma u^2 G(u)}{v^2}
-\pdv{}{v}\left( \frac{\Gamma  G(u) }{v} \right)
-\frac{2\Gamma G(u)}{v^2}
\right)F \right] \nonumber\\
&=& \frac{1}{2}\pdv{^2}{v^2} \left(B_{v} F \right)
-\pdv{}{v} \left( A_{v}F \right).
\label{eq:term-II}
\end{eqnarray}

Expanding the derivative of the $G(u)$ function using the definition from Eq.\ (\ref{eq:G}),
\begin{eqnarray}%
	\pdv{G(u)}{u} &=&  \frac{\mbox{erf}'(u)}{u^2} - \frac{\mbox{erf}(u)}{u^3} + \mbox{erf}'(u) =  -\frac{2}{u} G(u) + \mbox{erf}'(u) ,
\label{eq:dg}
\end{eqnarray}
and 
\begin{eqnarray}%
	\pdv{G(u)/v}{v} &=&  \frac{1}{v}\pdv{G(u)}{v} -\frac{G(u)}{v^2} =\frac{1}{v\sqrt{2}v_{th}}\pdv{G(u)}{u} -\frac{G(u)}{v^2} .
    \label{eq:dg/v}
\end{eqnarray}

\noindent Using  Eq.\ (\ref{eq:dg}) and Eq.\ (\ref{eq:dg/v}) $A_v$ from Eq.\ (\ref{eq:FP-2}) can be written as 
\begin{eqnarray}%
-A_v &=& \frac{2 \Gamma u^2 G(u)}{v^2}
-\pdv{}{v}\left( \frac{\Gamma  G(u) }{v} \right)
-\frac{2\Gamma G(u)}{v^2} = \frac{\Gamma}{v^2}\left( \mbox{erf}(u) - 2u\:\mbox{erf}'(u) + G(u) \right) . 
%
\end{eqnarray}

Expanding the pitch angle contribution, term III, from Eq.\ (\ref{eq:fp-original-g})
\begin{eqnarray}\label{eq:term-III}
III &=& \frac{\Gamma}{2v^3} \pdv{}{\mu} \left( (1 - \mu^2) \left( \text{erf}(u) - G(u) \right) \pdv{F}{\mu} \right) \nonumber  \\
&=& \pdv{}{\mu^2} \left( \frac{\Gamma}{2v^3}(1 - \mu^2) \left( \text{erf}(u) - G(u) \right) F \right) - \pdv{}{\mu} \left( \frac{\Gamma}{2v^3} \left( \text{erf}(u) - G(u) \right) \pdv{(1 - \mu^2)}{\mu}(1 - \mu^2) F\right) \nonumber  \\
&=& \frac{1}{2}\pdv{}{\mu^2} \left( \underbrace{\frac{\Gamma}{v^3}(1 - \mu^2) \left( \text{erf}(u) - G(u) \right)}_{B_{\mu}} F \right) - \pdv{}{\mu} \left( \underbrace{-\frac{ \Gamma \mu}{v^3} \left( \text{erf}(u) - G(u) \right)}_{A_\mu} F\right)  \nonumber  \\
&=& \pdv{}{\mu^2} \left( B_{\mu} F \right) - \pdv{}{\mu} \left( A_\mu F\right) .
\end{eqnarray}

Using It$\hat{\rm{o}}$'s lemma Eq.\ (\ref{spde}) can be written as an equivalent system of stochastic differential equations as
\begin{equation}
    d\mathbf{X} = \mathbf{A}(\mathbf{X},t) dt + \sqrt{\mathbf{B}} \ d\mathbf{W}(t) ,
   \label{itos}
\end{equation}
substituting Eqs.\ (\ref{eq:term-I}), (\ref{eq:term-II}) and (\ref{eq:term-III}) into Eq.\ (\ref{eq:fp-original-g}), and
using the definition from Eq.\ (\ref{itos}), we can write as
\begin{eqnarray*}
    dv &=&  (A_v+A_{vs}) dt + \sqrt{B_{v}+B_{vs}}dW(t) , \\
    d\mu &=&  A_\mu dt + \sqrt{B_{\mu}}dW(t) , \\
    dz &=&  A_z dt ,
\end{eqnarray*}
where the coefficients are defined as
\begin{align*}
A_{v} &= -\frac{\Gamma}{v^2}\left( \text{erf}(u) - 2\text{erf}'(u) + G(u) \right), & B_{v} &= \frac{2\Gamma G(u)}{v}, \\
A_{vs} &= \frac{2D_{\text{turb}}}{v} + \pdv{D_{\text{turb}}}{v}, & B_{vs} &= 2D_{\text{turb}}, \\
A_{\mu} &= -\frac{\Gamma \mu}{v^3} \left( \text{erf}(u) - G(u) \right), & B_{\mu} &= \frac{\Gamma}{v^3}(1 - \mu^2) \left( \text{erf}(u) - G(u) \right), \\
A_{z} &= \mu v. &
\end{align*}
Finally, it is important to note that, because the SDEs are solved for $v$ rather than $E$, any energy histogram should be weighted by $mv$ to properly account for the change of variables in the probability density function, since
\begin{equation}
\int_{E_1}^{E_2} F(E) dE = \int_{V(E_1)}^{V(E_2)} F(E(v)) \frac{dE}{dv}dv = \int_{V(E_1)}^{V(E_2)} F(v) mvdv .
\end{equation}

\section{Sigmoid Fitting} \label{appendix-fitting}
\begin{figure*}[h]
	\centering
	\includegraphics[width=1.0\textwidth]{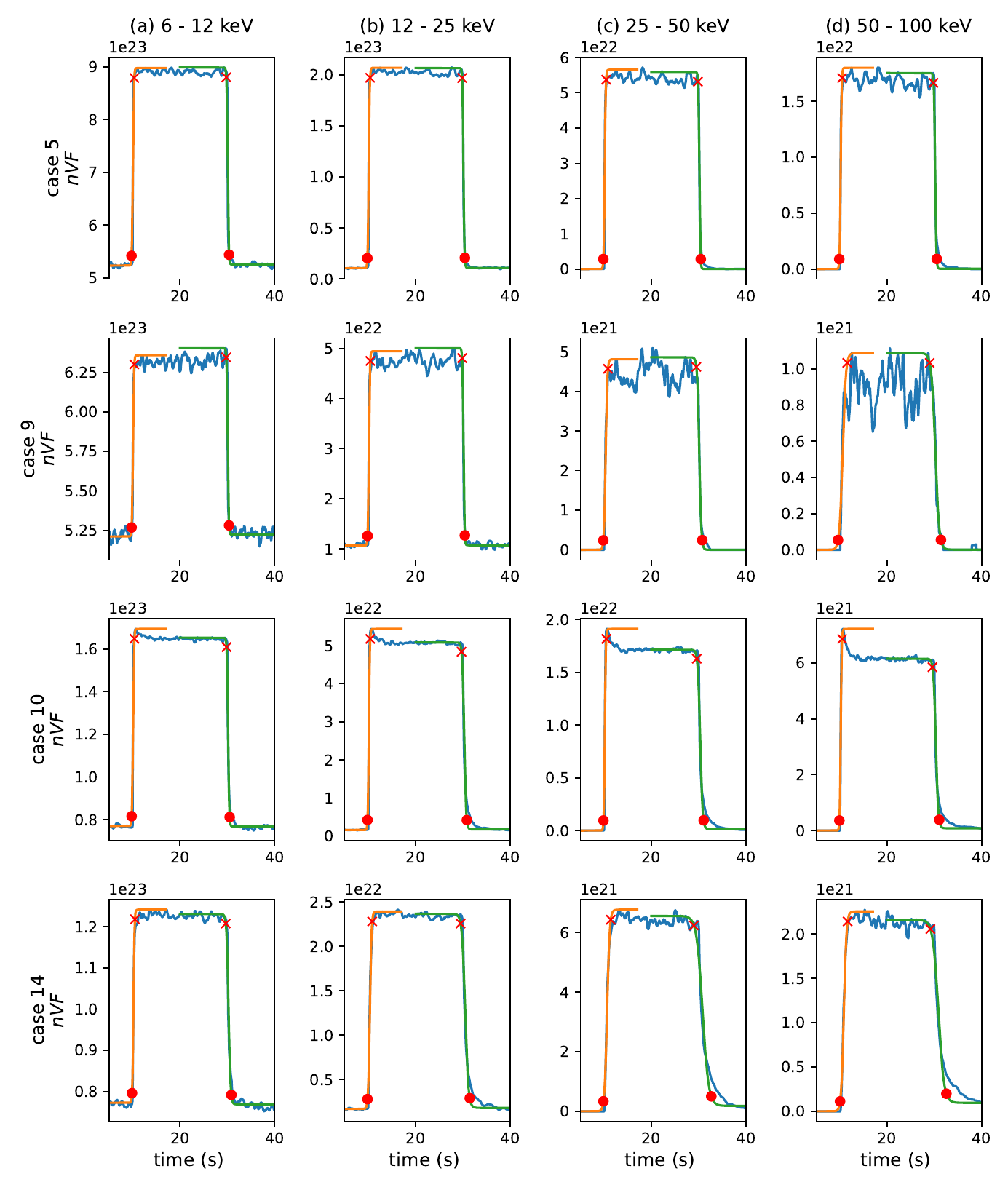}
	\caption{Blue lines shows $nVF$ from simulations described in Table \ref{tab:setup}, orange and green lines represent fittings using the sigmoid function, $S(t)$, is defined according to Eq.\ (\ref{eq:sigmoid}). Red dots represent where $S(t)=5\%$ and  red crosses where $S(t)=95\%$. The fitting was used to quantify the time lag shown in Fig.\ \ref{fig:hat-fitting}.}
	\label{fig:fitting}
\end{figure*}

To quantify the response time scale when a square pulse driver is applied (simulations 5 to 14), we fit the $nVF$ signal across different energy bands using the sigmoid function in Eq.~(\ref{eq:sigmoid}). The $nVF$ signal was normalized between 0 and 1, and the fit was used to extract $\tau_{\rm{on}}$ and $\tau_{\rm{off}}$. The response time scales $\tau_{\rm{on}}$ and $\tau_{\rm{off}}$ are defined as the time interval during which the sigmoid function transitions from $S(t)=5\%$ to $S(t)=95\%$. Figure~\ref{fig:fitting} shows the $nVF$ time series for multiple energy bands along with the corresponding sigmoid function fits.

\bibliography{references}{}
\bibliographystyle{aasjournal}



\end{document}